%% file: fluctuating_geometry.tex
\documentclass[12pt]{article}
\usepackage{hyperref}
\usepackage{graphicx}
\usepackage{amsmath}
\usepackage{amsthm}
\usepackage{amssymb}
\usepackage{tipa}
\usepackage{subfig} 
\usepackage{xcolor}
\usepackage{mathrsfs}
\theoremstyle{plain}

\usepackage{tikz}
\usepackage{pgfplots}
\usepackage{tikz-cd}
\usetikzlibrary{arrows}
\usetikzlibrary{intersections}
\usetikzlibrary{shapes.geometric}
\usetikzlibrary{decorations.pathmorphing, patterns,shapes}

\usepackage[nosort]{cite}

\newlength{\abstractwidth}
\renewcommand{\title}[1]{\vbox{\center\bf{\Large{#1}}}\vspace{5mm}}
\renewcommand{\author}[1]{\vbox{\center#1}\vspace{5mm}}
\newcommand{\address}[1]{\vbox{\center\em#1}}

\renewcommand{\bar}{\overline}
\renewcommand{\tilde}{\widetilde}

\renewcommand{\leq}{\leqslant}
\renewcommand{\geq}{\geqslant}

\newcommand{\nn}{\nonumber}

\newcommand{\calA}{\mathcal{A}}

\newcommand{\calC}{\mathcal{C}}

\input{XLdef}

\begin{document}

\begin{titlepage}
\begin{center}
\hfill \\
\hfill \\
\vskip 1cm

\title{Holographic coherent states from random tensor networks}

\author{Xiao-Liang Qi$^1$, Zhao Yang$^1$ and Yi-Zhuang You$^2$}
\address{
$^1$Department of Physics, Stanford University, Stanford, CA 94305, USA\\
$^2$Department of Physics, Harvard University, Cambridge, MA 02138, USA
}

\end{center}
  
\begin{abstract}
Random tensor networks provide useful models that incorporate various important features of holographic duality. A tensor network is usually defined for a fixed graph geometry specified by the connection of tensors. In this paper, we generalize the random tensor network approach to allow quantum superposition of different spatial geometries. We setup a framework in which all possible bulk spatial geometries, characterized by weighted adjacient matrices of all possible graphs, are mapped to the boundary Hilbert space and form an overcomplete basis of the boundary. We name such an overcomplete basis as holographic coherent states. A generic boundary state can be expanded in this basis, which describes the state as a superposition of different spatial geometries in the bulk. We discuss how to define distinct classical geometries and small fluctuations around them. We show that small fluctuations around classical geometries define ``code subspaces" which are mapped to the boundary Hilbert space isometrically with quantum error correction properties. In addition, we also show that the overlap between different geometries is suppressed exponentially as a function of the geometrical difference between the two geometries. The geometrical difference is measured in an area law fashion, which is a manifestation of the holographic nature of the states considered. 
\end{abstract}
\end{titlepage}

\tableofcontents

\baselineskip=17.63pt

\section{Introduction}

\begin{figure}[h]
\centerline{\includegraphics[width=2.5in]{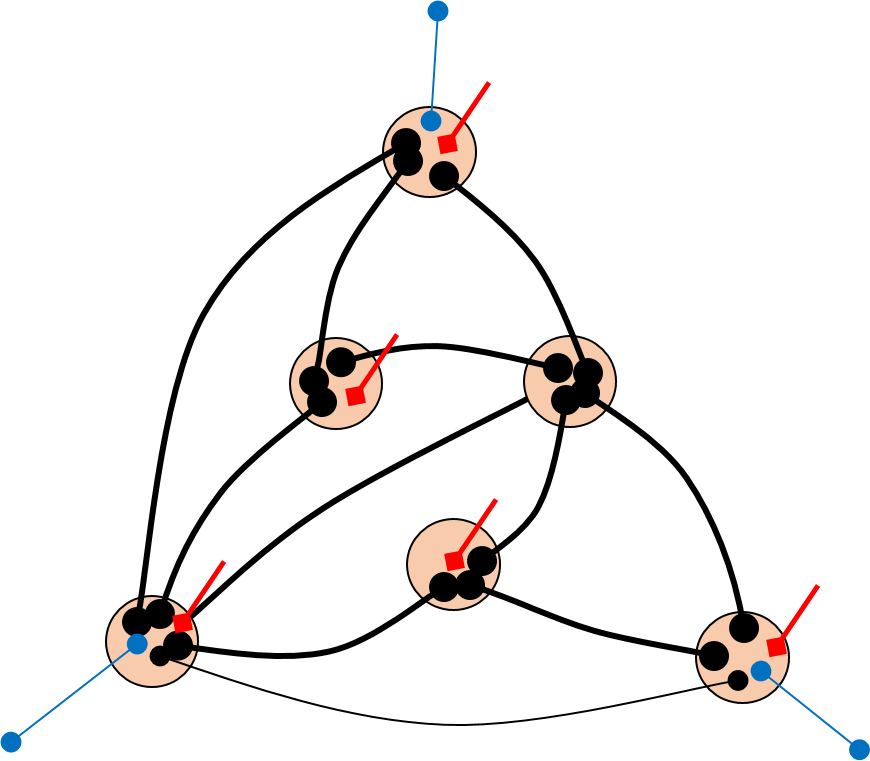}}
\caption{An example of random tensor network. }\label{fig:RTN}
\end{figure}

The holographic duality\cite{witten1998anti,gubser1998gauge,maldacena1999large} was proposed as a duality between quantum gravity in $d+1$ dimensions and quantum field theory in $d$ dimensions. The correspondence was originally proposed between the partition function and correlation functions of the two theories. The large $N$ limit of the boundary quantum field theory corresponds to the bulk semiclassical limit (the limit of small Newton constant $G_N$). The role of quantum entanglement in holographic duality was explicitly reflected by the Ryu-Takayanagi formula\cite{ryu2006holographic} and its generalizations\cite{hubeny2007covariant,faulkner2013quantum,dong2016deriving,dong2016gravity}, which relates entanglement entropy of a boundary region to area of extremal surfaces. The relation of entropy and area motivated the proposal that tensor networks may provide a ``microscopic" framework for understanding holographic duality\cite{swingle2012entanglement,swingle2012constructing}. Tensor networks, or projected entangled pair states (PEPS) is an approach to construct entangled quantum many-body states\cite{white1992density,
  white1993density, verstraete2004density,
  verstraete2004renormalization, vidal2008class,gu2009tensor}. For a graph (see Fig. \ref{fig:RTN}), the corresponding PEPS is obtained by first preparing an EPR pair for each link, and then projecting all qubits at the same vertex to a pure state specified by the tensor at that vertex. This procedure leads to a many-body state of the remaining qubits living at the end of dangling legs of the network. The advantage of the tensor network description is that the entanglement structure of the state is encoded explicitly in the geometry of the network. In particular, the entanglement entropy of each region $A$ is bounded by the minimal number of links that separate $A$ and its complement, multiplied by $\log D$ with $D$ the bond dimension of each tensor. This is the analog of RT formula. 
  
To use tensor networks to understand holographic duality, a key question is what states correspond to semiclassical bulk geometry. As we learn in holographic duality, such states must satisfy various conditions\cite{heemskerk2009holography,el2012emergent}, such as the negative tripartite mutual information\cite{hayden2013holographic}. Various tensor network models\cite{qi2013exact,pastawski2015holographic,yang2016bidirectional,hayden2016holographic} have been proposed to incorporate desired features of holographic duality. Among them, the random tensor networks\cite{hayden2016holographic} are shown to realize many features of holographic duality naturally, including the RT formula with quantum corrections, and the quantum error correction property of the bulk-boundary holographic mapping\cite{almheiri2015bulk}.  However, there are holographic properties that are not reproduced by tensor networks, such as the Renyi entropy behavior\cite{dong2016shape, dong2016gravity}. There are also obviously many other open questions that have not been addressed in the tensor network framework. 

Among the open questions, an essential one is how to describe quantum superposition of different geometries, as is required for a quantum gravity theory. This is also a necessary step towards understanding Einstein equation and graviton excitations in the bulk. In this paper we make a small progress along this direction by setting up a framework for describing quantum superposition of tensor network states on arbitrary geometries.  
We generalize the random tensor network approach in Ref. \cite{hayden2016holographic} and define a linear map between geometries in the bulk and quantum states on the boundary. A geometry is described by the adjacient matrix $a_{xy}$ of a weighted (unoriented) graph (with fixed number of vertices and arbitrary connectivity), which is defined as a basis vector $\ket{\ke{a_{xy}}}$ in the bulk. The linear map defined by random tensors then maps each such basis state to a quantum state $\ket{\Psi\kd{\ke{a_{xy}}}}$ on the boundary, which is the holographic state that is dual to this geometry. With this linear map it is straightforward to take superpositions between different geometries. We prove that for a fixed size of the boundary, a large enough number of bulk vertices make such a mapping an isometry from the boundary to the bulk. In other words, $\ket{\Psi\kd{\ke{a_{xy}}}}$ parameterized by the weighted adjacient matrix $a_{xy}$ form an overcomplete basis of the boundary, such that each state in the boundary Hilbert space can be mapped to a quantum superposition of different geometries. Due to the analog of boson coherent states (as will be elaborated more in later part of this paper), we name this basis of states ``holographic coherent states". 

Furthermore, this formalism allow us to consider small fluctuation around a classical geometry, and show that such small fluctuations for a ``code subspace"\cite{almheiri2015bulk} which is mapped to the boundary isometrically. (The precise meaning of ``classical geometry" and ``small fluctuation" will be given later. In short, a classical geometry means all nonzero entries of the weighted adjacient matrix $a_{xy}$ are large, while small fluctuations correspond to $a_{xy}\>a_{xy}+\delta a_{xy}$ with $\delta a_{xy}\ll a_{xy}$.
) Such small fluctuations can be considered as low energy states of the bulk quantum fields. The existence of bulk-boundary isometry in such subspaces guarantees that small fluctuations at different links of the graph are independent physical degrees of freedom. In other words, bulk locality emerges in such subspaces even if the whole bulk theory is intrinsically nonlocal. In addition, the bulk-boundary isometry satisfies the local reconstruction properties known in holographic duality. The structure of a boundary-to-bulk isometry in the whole boundary Hilbert space and a bulk-to-boundary isometry in code subspaces has been proposed as ``bidirectional holographic code" in Ref. \cite{yang2016bidirectional}, which is schematically summarized in Fig. \ref{fig:bidirectional}.

As an overcomplete basis, states $\ket{\Psi\kd{\ke{a_{xy}}}}$ for different geometry $a_{xy}$ do not correspond to orthogonal states of the boundary. However, we show that the overlap between different geometries are exponentially suppressed in the large $N$ limit. This is similar to ordinary boson coherent states that are used in mean-field approximation of superfluids and superconductors. Different coherent states are not orthogonal. But because their overlap is exponentially small, it is physically meaningful to consider them as physically different states, and therefore consider the condensate wavefunction as a physical order parameter field. An interesting difference of the geometrical states from ordinary coherent states is that the overlap between two states has a ``holographic" behavior. If two geometries $a_{xy}$ and $b_{xy}$ are distinct in a region $R$ and identical outside $R$, we prove that the overlap $\abs{\braket{\Psi\kd{\ke{a_{xy}}}}{\Psi\kd{\ke{b_{xy}}}}}$ is upper bounded by $e^{-c\abs{\gamma}}$ with $\gamma$ the area of a minimal surface bounding region $R$. $c$ is a constant determined by the entanglement entropy contributed by each link crossing the boundary. The area law form of the overlap is a manifestation of the fact that the states $\ket{\Psi\kd{\ke{a_{xy}}}}$ are consistent with the holographic principle---the fact that the physical degrees of freedom in a region $R$ are bounded by their area rather than volume. 

\addfig{3.3in}{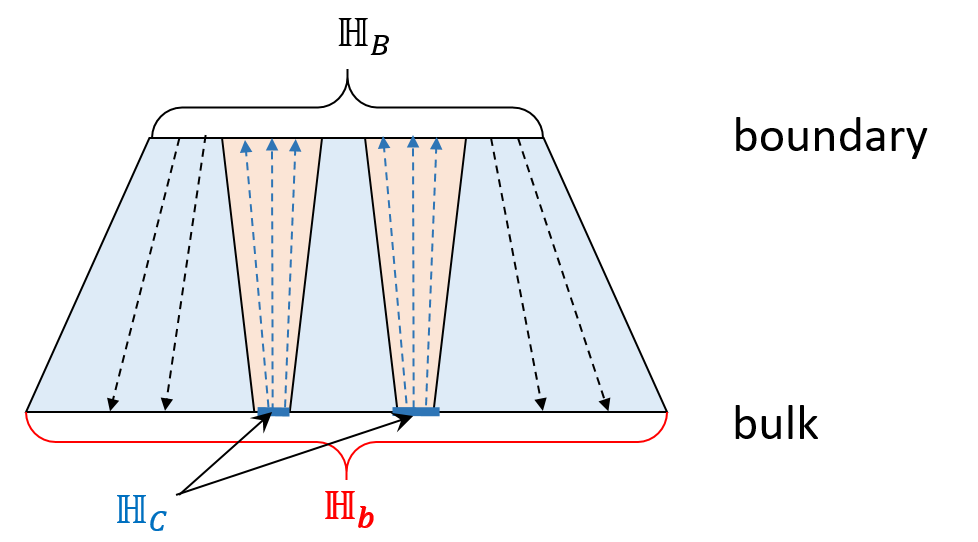}{Illustration of the structure of bidirectional holographic code defined by RTN. An isometry is defined from the boundary Hilbert space $\mathbb{H}_B$ to bulk Hilbert space $\mathbb{H}_b$ which maps each state in $\mathbb{H}_B$ to a superposition of geometries. In the code subspace $\mathbb{H}_C$ which consists of subspaces of small fluctuations around different classical geometries, an isometry is defined from bulk to boundary.\label{fig:bidirectional}}

The remainder of the article is organized as follows. In
Sec. \ref{sec:def} we present the general setup of our approach. In
Sec. \ref{sec:isometry} we study the condition of boundary-to-bulk
isometry. In Sec. \ref{sec:codesubspace} we investigate the definition
of classical geometries and the code subspaces with bulk-to-boundary
isometry. In Sec. \ref{sec:overlap} we study the overlap between
different classical geometries to show that distinct geometries are
almost orthogonal. Finally, the conclusion and further discussions are
given in Sec. \ref{sec:conclusion}.

\section{General framework}\label{sec:def}

We begin with a brief overview of the random tensor networks proposed in Ref. \cite{hayden2016holographic}. For a graph, such as the one in Fig. \ref{fig:RTN}, one first prepares a EPR pair of two qudits for each link, denoted by $\ket{xy}$. Then the RTN is defined by projecting all qudits on the site $x$ to a random pure state $\ket{V_x}$. If each qudit has dimension $D$, and site $x$ has $k$ neighbors, $\ket{V_x}$ is a random unit vector in a $D^k$-dimensional Hilbert space. The probability distribution of $\ket{V_x}$ is uniform, which means $\ket{V_x}$ and $U\ket{V_x}$ has the same probability for any unitary $U$. Alternatively, one can define $\ket{V_x}=U\ket{0}$ with $U$ a Haar random unitary operator and $\ket{0}$ a fixed reference state. For a graph $G$, the RTN state is expressed as
\bea
\ket{\Psi_G}=\prod_x\bra{V_x}\prod_{\avg{xy}\in G}\ket{xy}
\eea
with the $\avg{xy}\in G$ runs over (unoriented) edges in the graph $G$. 

From the definition of RTN, it is natural to see how to generalize this formalism to include superposition of different geometries (graphs)---The link state $\prod_{\avg{xy}\in G}\ket{xy}$ can be replaced by superpositions of such states on different graphs. To make this well-defined, one needs to modify the definition slightly to make sure the Hilbert space dimension of each vertex is identical for different graphs. This can be easily achieved by defining some auxiliary states on links that are absent in $G$. For each $\avg{xy}\notin G$, define a state $\ket{xy}_0=\ket{x}_0\ket{y}_0$ which is a direct product state and is orthogonal to $\ket{xy}$. Adding such direct product states do not change the entanglement structure of the system. Then if we replace $\prod_{\avg{xy}\in G}\ket{xy}$ by $\prod_{\avg{xy}\in G}\ket{xy}\prod_{\avg{xy}\notin G}\ket{xy}_0$, the dimension of each site is $D^{V-1}$ if the total number of vertices is $V$. Therefore the random states $\ket{V_x}$ can be chosen in a Hilbert space of dimension $D^{V-1}$ independent from $G$. Denote $\ket{P_G}= \prod_{\avg{xy}\in G}\ket{xy}\prod_{\avg{xy}\notin G}\ket{xy}_0$ as the ``parent state" before projection, then the superposition of two geometries $G,G'$ correspond to a boundary state
$a\ket{\Psi_G}+b\ket{\Psi_{G'}}=\prod_x\bra{V_x}\kd{a\ket{P_G}+b\ket{P_{G'}}}$. In other words, now we have a {\it linear map} between different graphs $G$ corresponding to different EPR pair configurations (in the {\it same} Hilbert space) to different boundary states. 

\addfig{3.5in}{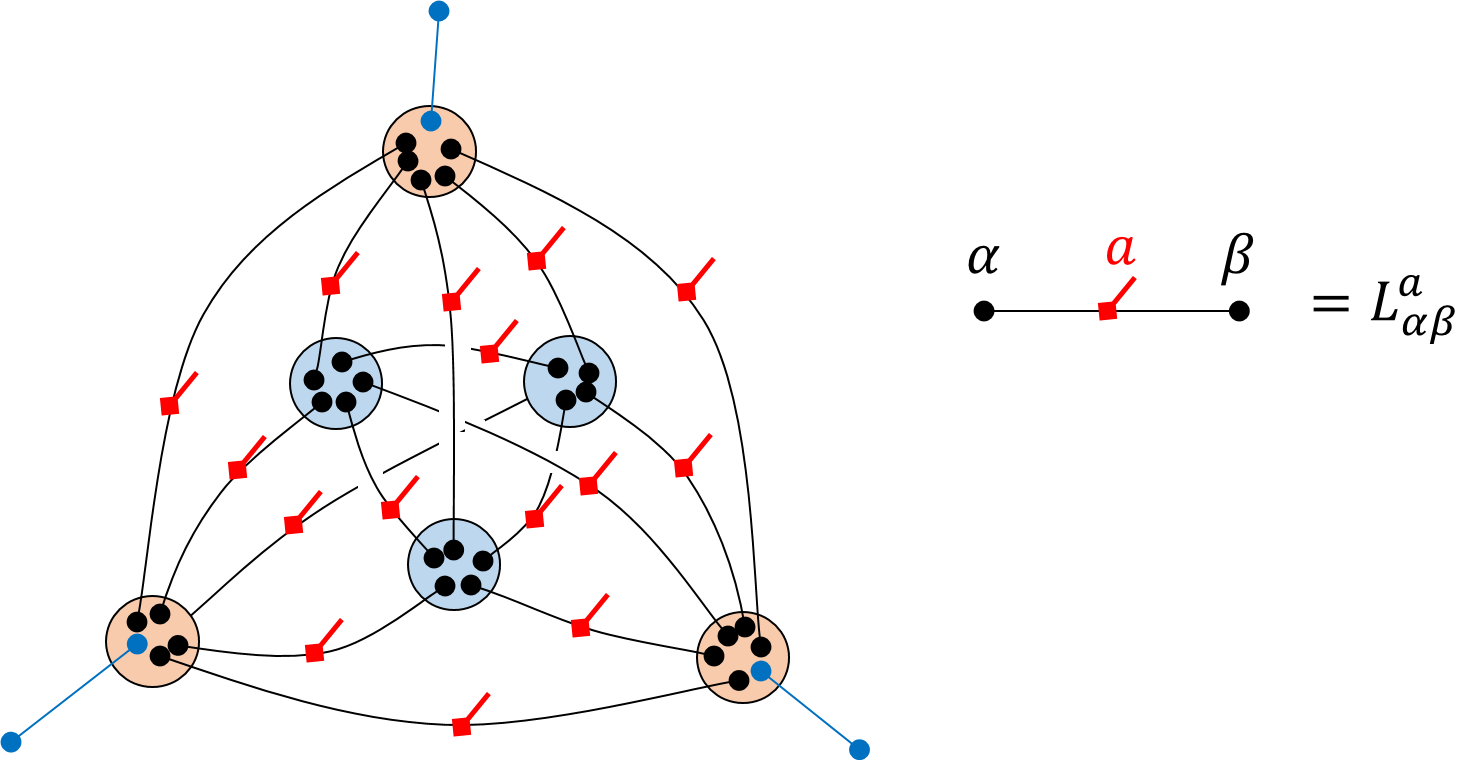}{Illustration of the random tensor network on a complete graph with link states. Each bulk link is a three-leg tensor $L_{\alpha\beta}^a$, and each vertex is a random tensor. The blue links are maximally entangled EPR pairs. The network defines a linear map between bulk link states (red lines) and boundary states (blue lines). \label{fig:fluctuating}}

Motivated by the discussion above, we consider a more general situation and define the following tensor network. Consider a complete graph with $V$ vertices, in which $V_B$ of them are labeled as ``boundary" vertices, and the rest of them $V_b=V-V_B$ are bulk vertices. For each pair of vertices $x,y$ ($x\neq y$), we define a three-leg tensor $L_a^{\alpha\beta}$ shown in Fig. \ref{fig:fluctuating}, with $a=0,1,2,...,D_L-1$ and $\alpha,\beta=1,2,...,D$.\footnote{Similar link variables have been introduced in perfect tensor networks in Ref. \cite{donnelly2016living} for a different but related purpose.} This tensor defines an isometry from index $a$ to indices $\alpha\beta$. In other words, states
\bea
\ket{a_{xy}}=L_a^{\alpha\beta}\ket{\alpha}_x\ket{\beta}_y
\eea
are orthonormal, {\it i.e.} $\braket{b_{xy}}{a_{xy}}=\delta_{ab}$. (Obviously this requires $D_L\leq D^2$.) The link variables $a_{xy}$ can be considered as specifying a weighted graph. Since we want the weight $a_{xy}$ to label entanglement in state $\ket{a_{xy}}$, we can require the entanglement entropy between $x$ and $y$ to be an increasing function of $a_{xy}$. For example, to be specific we can require 
\bea
S_x\kc{a_{xy}}=a_{xy}\log d,~\text{with~}a_{xy}=0,1,...,D_L-1,~d^{D_L-1}=D
\eea
which means $a_{xy}$ is the number of EPR pairs across the link, each with dimension $d$. The maximal $a_{xy}$ corresponds to a maximally entangled state. 

In addition, each boundary vertex is connected with a EPR pair state $\ket{xX}_B$ which entangles a qudit at vertex $x$ with one at the boundary physical site $X$. Then for each configuration $a_{xy}=0,1,2,...,D_L-1$, an RTN is defined by 
\bea
\ket{\Psi\kd{\ke{a_{xy}}}}=\prod_{x}\bra{V_x}\prod_{x\neq y}\ket{a_{xy}}\prod_x\ket{xX}_B\label{eq:fluctuatingRTN}
\eea
If we only want to incorporate superposition of RTN on different graphs, the simplest choice will be $D_L=2$, in which case a qubit at each link determines whether the link is connected (entangled) or not. However, it is more convenient to introduce a larger $D_L$, which makes it possible to define ``small" fluctuations around a classical geometry, as will be discussed in Sec. \ref{sec:codesubspace}.

The definition (\ref{eq:fluctuatingRTN}) can be considered as a linear map between the boundary Hilbert space (with dimension $D^{V_B}$) and the bulk Hilbert space spanned by the link qudits (with dimension $D_L^{V(V-1)/2}$). One can view this map as a holographic mapping that builds a correspondence between states (on the boundary) and geometries (in the bulk). It is straightforward to generalize the random average technique in Ref. \cite{hayden2016holographic} to the current setup, which is how we will investigate properties of this holographic mapping in the following sections.

\section{Boundary-to-bulk isometry}\label{sec:isometry}

In this section, we will study the holographic mapping from boundary to bulk, and show that it is an isometry under certain conditions. This result demonstrates that tensor network states $\ket{\Psi\kd{\ke{a_{xy}}}}$ for all configurations $\ke{a_{xy}}$ forms an overcomplete basis of the boundary Hilbert space, so that any boundary state can be expanded in this basis. 

The isometry condition requires 
\bea
\rho_B=\sum_{\ke{a_{xy}}}\ket{\Psi\kd{\ke{a_{xy}}}}\bra{\Psi\kd{\ke{a_{xy}}}}\propto \mathbb{I}\label{isometrycondition}
\eea
If we view the tensor network in Fig. \ref{fig:fluctuating} as an entangled state between boundary and bulk link qudits, the isometry condition is equivalent to the statement that the reduced density matrix $\rho_B$ is maximally mixed. To study $\rho_B$ we study its second Renyi entropy
\bea
e^{-S^{(2)}_B}=\frac{\trace{\rho_B^2}}{\trace{\rho_B}^2}
\eea
Similar to Ref. \cite{hayden2016holographic}, we study the random average of the numerator and denominator separately, and then study their fluctuations. When the fluctuation is small, we have $\overline{e^{-S^{(2)}_B}}\simeq {\overline{\trace{\rho_B^2}}}/{\overline{\trace{\rho_B}^2}}$. 

\subsection{The random-averaged isometry condition}

A commonly used trick in writing the Renyi entropy is to write
\bea
\trace{\rho_B^2}=\trace{X_B\rho_B\otimes \rho_B}=\trace{\kc{X_B\otimes I_{\overline{B}}}\kc{\rho_B\otimes \rho_B}}
\eea
with $\rho_B=\ket{\Psi\kd{\ke{a_{xy}}}}\bra{\Psi\kd{\ke{a_{xy}}}}$ the density matrix of the whole system, and $X_B$ the swap operator acting on two-copies of the system which permutes the two copies in $B$ region. More explicitly, if we denote an orthonormal basis of $B$ region as $\ket{n}_B$, then $X_B\ket{n}_B\otimes \ket{n'}_B=\ket{n'}_B\otimes \ket{n}_B$. 

For the state defined in Eq. (\ref{eq:fluctuatingRTN}), $\rho_B$ is
\bea
\rho_B&=&{\rm tr}_{b}\kd{\kc{\prod_x\ket{V_x}\bra{V_x}}\kc{\prod_{x\neq y}\rho_{xy}\otimes \prod_x\ket{xX}_B\bra{xX}_B}}\label{eq:rhoB}\\
\text{with~}\rho_{xy}&=&\frac1{D_L}\sum_{a=0}^{D_L-1}\ket{a_{xy}}\bra{a_{xy}}
\eea
Therefore
\bea
\overline{\trace{\rho_B^2}}&=&\trace{\kc{X_B\otimes\prod_x\overline{\ket{V_x}\bra{V_x}^{\otimes 2}}}\kc{\prod_{x\neq y}\rho_{xy}\otimes \prod_x\ket{xX}_B\bra{xX}_B}^{\otimes 2}}\nonumber\\
&=&C^{-1}\sum_{R\subseteq \text{bulk}}\trace{X_{B\cup R}\kc{\prod_{x\neq y}\rho_{xy}\otimes \prod_x\ket{xX}_B\bra{xX}_B}^{\otimes 2}}\label{eq:averagepurity}
\eea
Here we have used the mathematical fact that the random average $\overline{\ket{V_x}\bra{V_x}^{\otimes 2}}\propto \mathbb{I}_x\otimes \mathbb{I}_x+X_x$, with $X_x$ the swap operator defined in the same way as $X_B$, acting on all qudits at site $x$. The normalization constant $C=\kc{D^{2V-2}+D^{V-1}}^{V_b}\kc{D^{2V}+D^V}^{V_B}$.

The right-hand side of Eq. (\ref{eq:averagepurity}) is a sum over the purity of the state $\prod_{x\neq y}\rho_{xy}\otimes \prod_x\ket{xX}_B\bra{xX}_B$ for different regions $B\cup R$, with $R$ running over all $2^V$ subsets of the $V$ vertices. Since this state is simple, with only bipartite entanglement between different sites, the purity can be explicitly computed. In the same way as in Ref. \cite{hayden2016holographic}, the sum can be expressed as a partition function of a classical Ising model, with an Ising spin $s_x=\pm 1$ defined on each site. Each spin configuration corresponds to a region $R_\downarrow$ which is defined as the spin $s=-1$ domain. The action of the Ising model $\mathcal{A}\kd{\ke{s_x}}$ is defined such that $e^{- \mathcal{A}\kd{\ke{s_x}}}=\trace{X_{B\cup R_\downarrow }\kc{\prod_{x\neq y}\rho_{xy}\otimes \prod_x\ket{xX}_B\bra{xX}_B}^{\otimes 2}}$. Since the state on the righthand side only contains bipartite entanglement, the Ising model action only contains one-body and two-body terms:
\bea
\mathcal{A}\kd{\ke{s_x}}&=&-\frac J2\sum_{xy}\kc{s_x s_y-1}-\frac h2\sum_x s_x+\frac12\log D\sum_{x\in B} s_x\label{Isingaction}\\
\text{with~}h&=&\frac{V-1}2\log D_L,~J=s_b-\frac12\log D_L=\frac12\kc{S^{(2)}_x+S^{(2)}_y-S^{(2)}_{xy}}\nn
\eea
Here $0<s_b\leq \log D$ is the second Renyi entropy of site $x$ in the state $\rho_{xy}$, {\it i.e.} $e^{-s_b}={\rm tr}_x{\kc{{\rm tr}_y\rho_{xy}}^2}$, and the Ising coupling $J$ is half of the second Renyi mutual information between sites $x,y$ in the mixed state $\rho_{xy}$. The last term in the action sums over the $V_B$ boundary sites. 

The Ising model problem is simpler than that for a generic RTN in Ref. \cite{hayden2016holographic} because all pairs of $x,y$ are coupled equally. Consequently, all $V_b$ bulk vertices are equivalent, and all $V_B$ boundary vertices are equivalent. The action is therefore only a function of two integers, the number of down spins in the bulk vertices $n_b\in[0,V_b]$, and the number of down spins in the boundary vertices $n_B\in[0,V_B]$. 
\bea
\mathcal{A}\kd{\ke{s_x}}&=&\mathcal{A}(n_b,n_B)=\mathcal{A}_1+\mathcal{A}_2+\mathcal{A}_3\nn\\
\mathcal{A}_1&=&\log D_L (n_b+n_B)(n_b+n_B-1)/2,\nn\\
\mathcal{A}_2&=&s_b \kc{n_b+n_B}\kc{V-n_b-n_B},~\mathcal{A}_3=\log D \kc{V_B-n_B}\label{IsingactionnbnB}
\eea
The three terms $\mathcal{A}_{1,2,3}$ are contributions of links within region $R$, links between $R$ and its complement, and links from $\overline{R}$ to the boundary, respectively, as is illustrated in Fig. \ref{fig:renyi2}. 

\addfig{2.5in}{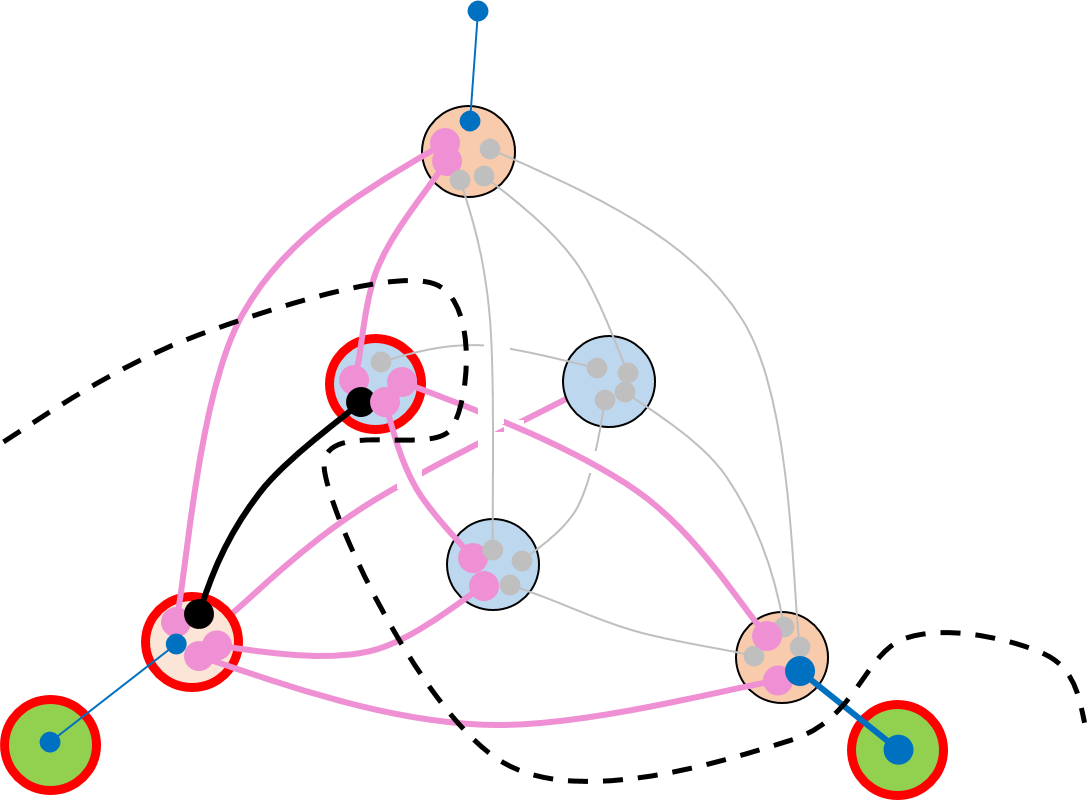}{Illustration of a spin configuration with $n_b=n_B=1$. The spins are $-1$ for all sites with a thick red circle, and $+1$ elsewhere. The dashed line is the domain wall across which the spin changes sign. The contribution to the action comes from three kinds of links, those within the spin down region (black thick line), those between opposite spins (pink thick line) and those connecting the spin up boundary sites to the boundary (blue thick line). These three contributions correspond to $\mathcal{A}_{1,2,3}$ in Eq. (\ref{IsingactionnbnB}) respectively. \label{fig:renyi2}}

With the action $\mathcal{A}(n_b,n_B)$, Eq. (\ref{eq:averagepurity}) becomes
\bea
\overline{\trace{\rho_B^2}}=C^{-1}\sum_{n_b=0}^{V_b}\sum_{n_B=0}^{V_B}\vect{V_b\\n_b}\vect{V_B\\n_B}e^{-\mathcal{A}\kc{n_b,n_B}}
\eea
For large $V_b, V_B$, this sum is dominated by the biggest term, which corresponds to the minimum of $\mathcal{S}\kc{n_b,n_B}=\mathcal{A}\kc{n_b,n_B}-\log\vect{V_b\\n_b}-\log\vect{V_B\\n_B}$. One can show that $\mathcal{S}\kc{n_b,n_B}$ reaches its minimum in the large $V_b,V_B$ limit  at one of the corners in region $n_b\in[0,V_b],~n_B\in[0,V_B]$. A detailed explanation can be found in Appendix.\ref{GR}. 
The same analysis applies to the denominator $\overline{\trace{\rho_B}^2}$, and the only difference is in the boundary term $\mathcal{A}_3$. 
\bea
\overline{\trace{\rho_B}^2}=C^{-1}\sum_{n_b=0}^{V_b}\sum_{n_B=0}^{V_B}\vect{V_b\\n_b}\vect{V_B\\n_B}e^{-\mathcal{\tilde{A}}\kc{n_b,n_B}}
\eea
with $\tilde{\mathcal{A}}=\mathcal{A}_1+\mathcal{A}_2+\tilde{\mathcal{A}}_3$ and $\tilde{\mathcal{A}}_3=n_B\log D $.

The isometry condition is satisfied if the dominant configuration for both the numerator and the denominator is given by $n_B=n_b=0$, which requires
\bea
\log D_L\frac{V(V-1)}2&>&V_B\log D\label{condition1}\\
\log D_L\frac{V_B(V_B-1)}2+s_bV_bV_B&>&V_B\log D\label{condition2}
\eea
Condition (\ref{condition1}) is simply a requirement that the bulk Hilbert space dimension $D_L^{V(V-1)/2}$ is larger than that of the boundary ($D^{V_B}$). Condition (\ref{condition2}) requires that the link state $\rho_{xy}$ is sufficiently entangled. In term of coupling $J=s_b-\frac12\log D_L$, the condition (\ref{condition2}) requires 
\bea
J>\frac1{V_b}\kc{\log D-\frac{V-1}2\log D_L}\label{condition2p}
\eea

Condition (\ref{condition1}) and (\ref{condition2}) are easy to satisfy. If we take the limit $V_b,V_B\rightarrow \infty$ with the ratio $V_B/V$ fixed, and keep $D,D_L$ finite, all conditions will be trivially satisfied. 

The isometry condition (\ref{isometrycondition}) allows an expansion of an arbitrary boundary state $\ket{\Phi}$ in this basis: $\ket{\Phi}=\sum_{\ke{a_{xy}}}\ket{\Psi\kd{\ke{a_{xy}}}}\braket{\Psi\kd{\ke{a_{xy}}}}{\Phi}=\sum_{\ke{a_{xy}}}\phi\kd{\ke{a_{xy}}}\ket{\Psi\kd{\ke{a_{xy}}}}$. This wavefunction is the analog of Wheeler-de Witt wavefunction\cite{dewitt1967quantum} of quantum gravity, although here we are only taking superpositions of spatial geometries.

\subsection{Fluctuations}

As we discussed earlier, the calculation of $\overline{\trace{\rho_B^2}}$ only tells us the average of second Renyi entropy if the fluctuation is small. The fluctuation can be studied by computing $\overline{\kc{\trace{\rho_B^2}^2}}-\left(\overline{\trace{\rho_B^2}}\right)^2$. As has been shown in Ref. \cite{hayden2016holographic}, the random average of a quantity like $\overline{\kc{\trace{\rho_B^2}^2}}$, which is quartic in $\rho_B$, can be expressed as a partition function of a statistical model with a pseudo-spin $g_x$ at each site taking values in the 4-element permutation group $S_4$. In general, any quantity in the form of $\trace{\overline{\rho^{\otimes k}}O_k}$, with operator $O_k$ acting on $k$ copies of the system,\footnote{For example, $\trace{\rho_A^{k}}$ which determines the $k$-th Renyi entropy can be written as $\trace{\rho^{\otimes k}C_{Ak}}$ with $C_{Ak}$ the cyclic permutation of the $k$ copies of systems in $A$ region.} is mapped to a partition function of a model with pseudo-spins in $k$-element permutation group $S_k$. Similar to the Ising model analyzed above, the statistical models for higher $k$ is also defined on a complete graph, which simplifies the problem. In Appendix \ref{app:fluctuation} we analyze these pseudospin models and obtain sufficient conditions for fluctuations such as $\overline{\kc{\trace{\rho_B^2}^2}}-\overline{\trace{\rho_B^2}}^2$ to be controlled. For bounding the fluctuation of the second Renyi entropy calculation, the sufficient conditions are the following: 
\bea
(V-1)&\gg& \frac{2\log D}{\log D_L}\label{sufficientcond1}\\
\abs{\trace{\rho_{xy}^{\otimes k}g\otimes h}}^2&< &\trace{\rho_{xy}^{\otimes k}g\otimes g}\trace{\rho_{xy}^{\otimes k}h\otimes h},~\forall g\neq h\in S_k\label{sufficientcond2}
\eea
with $k=4$ in the second equation. More details of the derivation will be given in Appendix \ref{app:fluctuation}. It is not difficult to see that conditions (\ref{sufficientcond1}) and (\ref{sufficientcond2}) imply the conditions we obtain earlier in Eq. (\ref{condition1}) (\ref{condition2p}). Condition (\ref{sufficientcond1}) can be easily satisfied in large volume $V$. Condition (\ref{sufficientcond2}) imposes addition constraints to the choice of states $\ket{a_{xy}}$ and $\rho_{xy}$, but is also not hard to satisfy, as we will discuss in more details in Appendix \ref{app:fluctuation}. We also give an explicit example of $\ket{a_{xy}}$  in Appendix \ref{app:spinexample} which satisfies condition (\ref{sufficientcond2}) for general $k$. 



\section{Bulk-to-boundary isometry in code subspaces}\label{sec:codesubspace}

Since the bulk basis $\ket{\Psi\kd{\ke{a_{xy}}}}$ is generically
overcomplete, the mapping from bulk to boundary defined by our random
tensor network is not injective. However, holographic duality requires
that small fluctuations around a classical geometry are independent physical states on the boundary. For example, if we consider a
dilute gas of gravitons in the bulk, the total degree of freedom of
the gas is proportional to volume. Gravitons at different bulk
locations should be dual to independent degrees of freedom on the
boundary, since graviton creation/annhilation operators should be
mapped to independent operators on the boundary by the dictionary of
holographic duality. This requirement means that there should be a
bulk-to-boundary isometry in the subspace of such small
fluctuations. 
The bulk small fluctuations are mapped to a subspace of the boundary
Hilbert space, named as the ``code
subspace"\cite{almheiri2015bulk,pastawski2015holographic}. Each
geometry corresponding to a configuration $a=\ke{a_{xy}}$ defines a
code subspace $\mathbb{H}_C[a]$. The mapping of such small
fluctuations to the boundary should satisfy the following local
reconstruction property: Each region on the boundary $A$ corresponds
to a minimal surface $\gamma_A$ in the bulk that is homologous to
it. The region enclosed by $A\cup \gamma_A$ is the entanglement wedge
$E_A$\footnote{More precisely, $E_A$ here is the intersection of the
  entanglement wedge and the spatial slice. Since we will always be
  dealing with a spatial slice, we neglect this difference and call
  $E_A$ the entanglement wedge.}. A bulk operator acting in the
subspace of small fluctuations (the {\it code subspace}) in the bulk
region $E_A$ can be reconstructed in boundary region $A$. Since each
bulk point can be enclosed by the entanglement wedges of different
boundary regions, information in the bulk can be recovered from
different boundary regions, making the bulk-boundary map in the code
subspace a quantum error correction code\cite{almheiri2015bulk}. The
bulk-boundary isometry and local reconstruction is illustrated in
Fig. \ref{fig:localreconstruction}.

\addfig{4in}{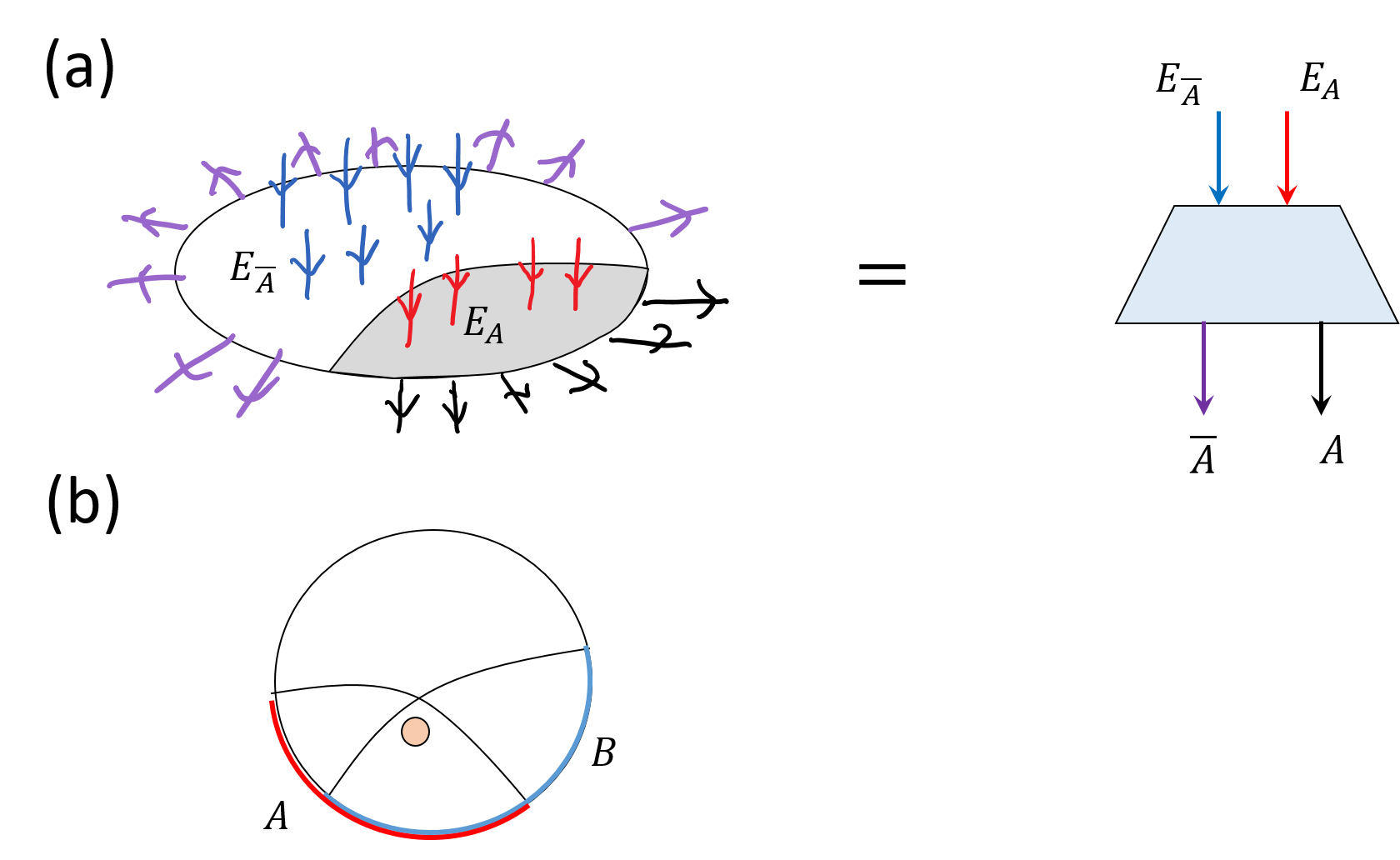}{\label{fig:localreconstruction} (a) Illustration of the bulk-boundary map defined in the code subspace. The mapping from the whole bulk subspace to the boundary is an isometry. Furthermore, the local reconstruction property requires that degrees of freedom in $E_{\overline{A}}$ which is the entanglement wedge of $A$ can be reconstructed in $A$, which means an isometry is defined from $E_A$ to $A$ for arbitrary states in $E_{\overline{A}}$ and $\overline{A}$. (b) A small region in the bulk (orange disk) can be reconstructed in different boundary regions such as $A,B$. }

In the following we will explain how our formalism of fluctuating geometry allows the definition of small fluctuations and code subspaces. We first define classical geometries and small fluctuations in our setup, and then study the ovelaps between different classical geometries to verify that macroscopically different geometries indeed correspond to almost orthogonal states on the boundary.

\subsection{Classical geometry and the code subspace}\label{sec:classicalgeo}

Each configuration $\ke{a_{xy}}$ corresponds to a ``geometry" ({\it i.e.} a weighted graph), but if $a_{xy}$ takes arbitrary values, one cannot define what fluctuations are considered ``small". With a large link variable dimension $D_L$, one can define a classical geometry as one with all nontrivial links ($a_{xy}\neq 0$) contributing a large entropy $\propto D_L$, and then define small fluctuations as fluctuations of $a_{xy}$ that are small compared to $D_L$. 

For concreteness, 
we pick a value of link variable $a_0$ with $0<\frac{a_0}{D_L-1}<1$, and take the limit $D_L,D\rightarrow \infty$ with $\frac{a_0}{D_L-1}$ fixed. We define a classical geometry by a state $\ket{\Psi\kd{\ke{a_{xy}}}}$ with all $a_{xy}$ equal to either $a_0$ or $0$. \footnote{It is straightforward to generalize the following discussion to states with different $a_0$ on different links as long as all of them are taken to infinity with the ratio $a_0/(D_L-1)$ fixed.} For such states, we can define an adjacient matrix $K$ with $K_{xy}=0,1$, such that $a_{xy}=K_{xy}a_0$. 

Now define a range of small fluctuation $\Lambda\ll a_0$. In the limit $D_L\>\infty$, $\Lambda$ is kept finite. Then we define small fluctuations around the classical geometry $K_{xy}a_0$ as all states $\ket{\Psi\kd{\ke{a_{xy}}}}$ satisfying
\bea
\elist{a_{xy}\in[a_0-\Lambda,a_0+\Lambda],&~\text{if~}K_{xy}=1\\
a_{xy}\in[0,2\Lambda],&~\text{if~}K_{xy}=0}
\eea
This range of $a_{xy}$ defines a subspace of the bulk, which is mapped to the boundary by the random tensor network. The definition of the classical geometry and small fluctuation subspace is illustrated in Fig. \ref{fig:semiclassical}. 

\addfig{4in}{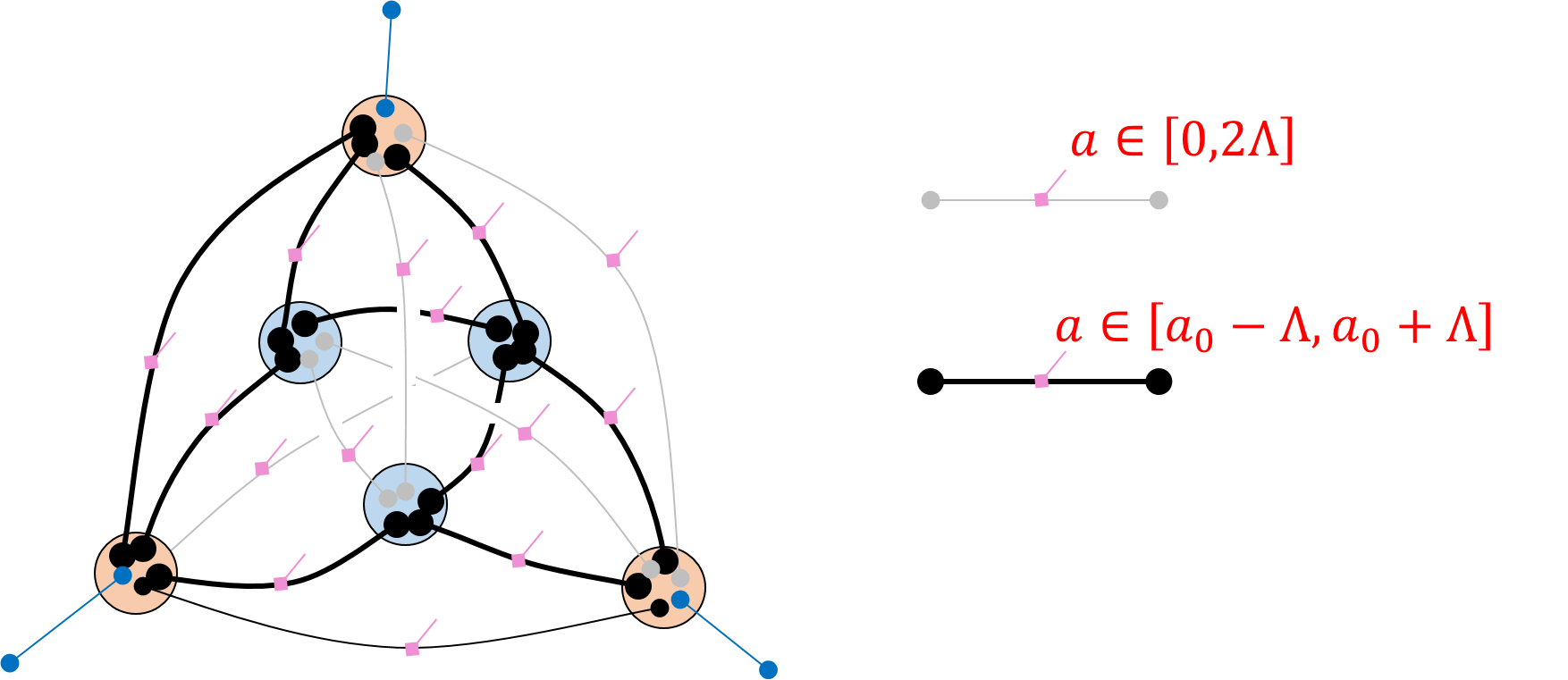}{\label{fig:semiclassical} Illustration of small fluctuations around a classical geometry. In the classical geometry, the black thick lines and grey thin lines are connected links with $a_{xy}=a_0$ and disconnected links with $a_{xy}=0$, respectively. The fluctuations are encoded by fluctuation of link quantum number $a$ around the classical value in a small range. }

To study whether the bulk-boundary map is an isometry, we carry the same calculation as in Sec. \ref{sec:isometry} to evaluate the second Renyi entanglement entropy between bulk and boundary. An isometry is defined if the bulk subspace is maximally entangled with the boundary. The calculation is exactly parallel to that in Sec. \ref{sec:isometry}, except that the bulk link state $\rho_{xy}$ in Eq. (\ref{eq:rhoB}) is replaced by
\bea
\rho_{xy}=\elist{\rho_1=\frac1{2\Lambda+1}\sum_{\delta a_{xy}=-\Lambda}^{\Lambda}\ket{a_0+\delta a_{xy}}\bra{a_0+\delta a_{xy}},&~\text{if~}K_{xy}=1\\
\rho_2=\frac1{2\Lambda+1}\sum_{\delta a_{xy}=0}^{2\Lambda}\ket{\delta a_{xy}}\bra{\delta a_{xy}},&~\text{if~}K_{xy}=0}\label{rhobulkCode}
\eea
The Ising action is changed correspondingly to
\bea
\mathcal{A}\kd{\ke{s_x}}&=&\mathcal{A}_0\kd{\ke{s_x}}+\delta\mathcal{A}\kd{\ke{s_x}}\nn\\
\mathcal{A}_0\kd{\ke{s_x}}&=&-\frac{J_1}2\sum_{\avg{xy}\in K}\kc{s_xs_y-1}+\frac12\log D\sum_{x\in B} s_x\nn\\
\delta \mathcal{A}\kd{\ke{s_x}}&=&-\frac{J_2}2\sum_{\avg{xy}\neq K}\kc{s_xs_y-1}-\frac{h_C}2\sum_x s_x\label{IsingactionCode}\\
\text{with~}h_C&=&\frac{V-1}2\log\kc{2\Lambda +1},\nn\\
J_1&=&s_b\kd{\rho_1}-\frac12\log\kc{2\Lambda+1},~J_2=s_b\kd{\rho_2}-\frac12\log\kc{2\Lambda+1}\nn
\eea
Here $J_{1,2}$ are half the Renyi mutual information of strong and weak link states $\rho_{1,2}$ in Eq. (\ref{rhobulkCode}), respectively. 
In the limit of $J_1\gg J_2$, $\log D\rightarrow \infty$ with $h_C$ and $\Lambda$ finite, $\mathcal{A}_0$ is the leading term in the action, and
$\delta\mathcal{A}$ is a subleading correction.

The analysis of this action is essentially the same as the original RTN case in Ref. \cite{hayden2016holographic}. The boundary term prefers $s_x=-1$, while the bulk pinning field $h_C$ prefers $s_x=+1$. Isometry condition is satisfied in the limit $\log D\>\infty,~J_1\>\infty$ if the lowest action configuration is $s_x=-1$ everywhere, which corresponds to an entropy $h_CV=\frac{V(V-1)}2\log\kc{2\Lambda +1}=\log \kc{{\rm dim}\mathbb{H}_C}$. In order for this configuration to have the lowest action, one requires that creating any spin up domain $R$ costs a positive action. Denoting the action of a spin configuration with $s_x=+1$ in $R$ and $s_x=-1$ elsewhere as $\mathcal{A}_R$, the isometry requirement is
\bea
\mathcal{A}_R-\mathcal{A}_\emptyset&=&\kc{J_1-J_2}\abs{\pa R}+J_2\abs{R}\kc{V-\abs{R}}\nn\\
& &+\log D\abs{R\cap B}-h_C|R|>0,~\forall R\subseteq \text{bulk}\label{conditionCode}
\eea
where the first two terms are action cost from the two-body interaction terms, the third term is the action cost from boundary pinning fields, while the last term is the action saved by the external field term $h_C$. $\abs{R}$ is the number of vertices in $R$ and $\abs{\pa R}$ is the number of links connecting $R$ and its complement in graph $K$ (excluding the boundary links). $\abs{\pa R\cap B}$ is the number of links connecting $R$ with boundary, {\it i.e.} the number of boundary sites in $R$.  

For sufficiently large $J_1,~\log D$ and finite $J_2,~h_C$, condition (\ref{conditionCode}) is satisfied. To obtain a more explicit understanding on the requirements, in the following we derive a sufficient condition which guarantees that the isometry condition (\ref{conditionCode}) is satisfied for {\it all} classical geometries. Denote $N$ and $M$ as the number of interior sites and boundary sites in $R$, respectively, such that $\abs{R}=N+M$ and $\abs{R\cap B}=M$. The action cost $\Delta \mathcal{A}\equiv \mathcal{A}_R-\mathcal{A}_\emptyset$ is a function of $N,M$ and the graph dependent parameter $\abs{\pa R}$. If we are considering a particular given graph, $\abs{\pa R}$ is not independent from $N$ and $M$. However, simplification occurs when we require condition (\ref{conditionCode}) to hold for all $R$ and for {\it all} classical geometries. By varying the graph, one can always vary $\abs{\pa R}$ of a given region $R$ in the range $\kd{0,\frac{(N+M)(V-N-M)}2}$). Therefore we can view the action cost $\Delta\mathcal{A}$ as a function of three independent variables $N,M,\abs{\pa R}$. This simplified the problem of minimizing $\Delta\mathcal{A}$, because the function in Eq. (\ref{conditionCode}) does not have local minimum in term of $N,M$ and $\abs{\pa R}$. Thus the minimum can only occur at corners of the parameter space. Given that $J_1-J_2>0$, the minimum always occurs at $\abs{\pa R}=0$, in which case $\Delta \mathcal{A}=J_2\kc{N+M}\kc{V-N-M}+M\log D-h_C(N+M)$. Evaluating $\Delta \mathcal{A}$ at the four corners $N=0$ or $V_b$ and $M=0$ or $V_B$ leads to two nontrivial conditions:
\bea
\Delta\mathcal{A}(V_b,V_B)>0&\=>&V_B \log D > \frac{V(V-1)}{2} \log (2\Lambda +1)\\
\Delta\mathcal{A}(V_b,0)>0&\=>&J_2V_bV_B>\frac{V_b(V-1)}2\log\kc{2\Lambda+1}
\eea
In summary the two sufficient conditions are
\bea
\log D &>& \frac{V(V-1)}{2 V_B} \log (2\Lambda +1)\label{codeSScond1}\\
J_2 &>&\frac{(V-1)}{2 V_B} \log (2\Lambda +1)\label{codeSScond2}
\eea
Physically, the first condition (\ref{codeSScond1}) is simply the requirement that the bulk code subspace has smaller dimension than the boundary. The second condition requires that even weak links with coupling $J_2$ provide strong enough entanglement to propagate information from bulk to boundary isometrically.
It should be noted that condition (\ref{codeSScond1}) requires $D$ to
grow exponentially with volume $V$ (if we fix the ratio $V_B/V$). This
is necessary since the bulk code subspace dimension grows with
$(2\Lambda+1)^{V(V-1)/2}$. Besides, Eq.\ref{codeSScond2} only
  requires $J_2$ to be a $O(1)$ number in this limit. If we consider
a limit $V\>\infty$ with large but finite $D$, it will be impossible
to faithfully represent all link variable fluctuations $\delta a_{xy}$
to the boundary. However, it is probably still possible to define a
code subspace with lower bound dimension, which contains bulk
excitations with a low enough density. (An example of such kind of
code subspace was discussed in Ref. \cite{yang2016bidirectional}.)
Such a code subspace which is not a direct product of Hilbert spaces
of each link is probably closer to the code subspace in AdS/CFT,
consisting low energy bulk quantum field theory excitations.






\subsection{Local reconstruction properties}\label{sec:localreconstruction}

Now we further investigate the local reconstruction properties of the bulk-boundary isometry. The local reconstruction requirement can be phrased in an entanglement entropy calculation. In the old setup of tensor networks with fixed geometry, shown in Fig. \ref{fig:localreconstruction}, one can view the bulk-boundary map as a quantum state that contains four partitions $A,\overline{A},E_A,E_{\overline{A}}$. The requirement that $A$ contains all information about $E_{A}$ is equivalent to the statement $I(E_A:\overline{A})=S\kc{E_A}+S(\overline{A})-S(E_A\overline{A})=0$. In the following we will evaluate the second Renyi entropy version of the mutual information. In the large $D$ limit when the fluctuation of Renyi entropies are small, we expect the von Neumann entropy to be equal to the Renyi entropy. Before proceeding, we would like to note that in the current setup the bulk degrees of freedom are defined on links, so that the bulk Hilbert space do not factorize into different regions. For a boundary region $A$, one can still define an entanglement wedge $E_A$, such that all edges with both ends contained in $E_A$ can be reconstructed from $A$. This is illustrated in Fig. \ref{fig:code}.

\addfig{3in}{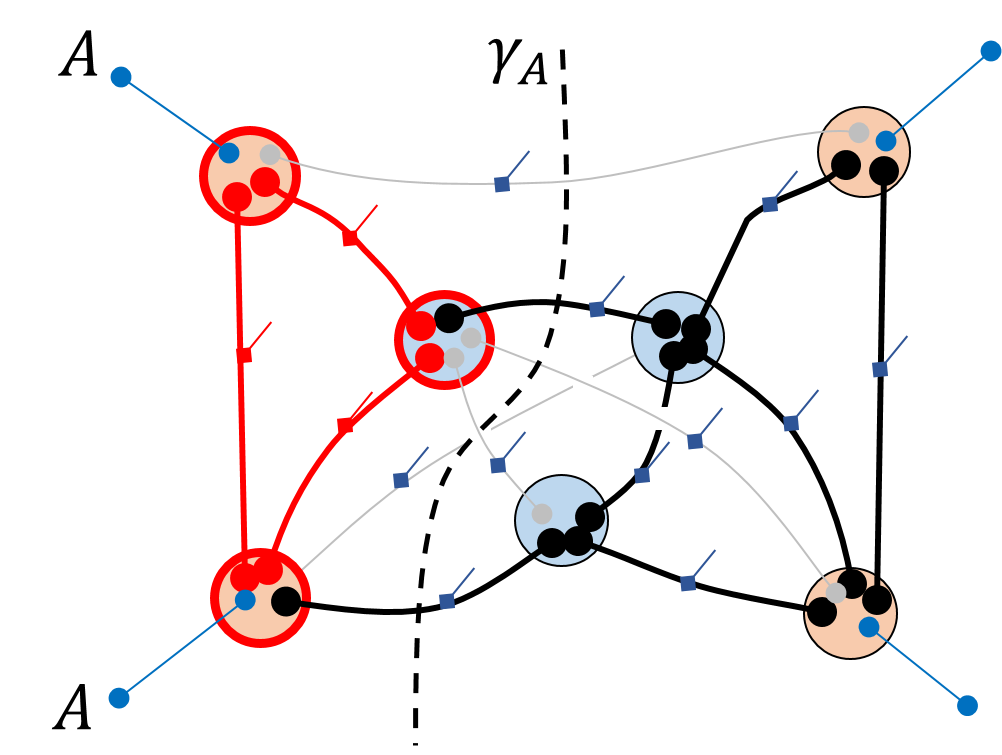}{Illustration of the entanglement wedge of a boundary region $A$. The vertices with red circles are the entanglement wedge $E_A$, enclosed by $A$ and the minimal surface $\gamma_A$. The code subspace that can be locally reconstructed in region $A$ are labeled by links with both ends in $E_A$, marked by red bulk lines. (For clarity we have only drawn a few of the unconnected (grey) links. ) \label{fig:code}}

Since we need to compute Renyi entropy of regions including both bulk and boundary, we should not trace over bulk link variables to obtain a reduced density matrix $\rho_{xy}$. Instead we treat the whole RTN with bulk and boundary indices as a state, and map the second Renyi entropy calculation to an Ising model partition function. All dangling ends of the tensor network in bulk and boundary correspond to fixed external spins that couple to the dynamical Ising spins defined on bulk vertices. We denote the dynamical Ising spins as $s_x$, and the external spins as $m_X$ on the boundary and $m_{xy}$ on bulk links. $s_x,m_X,m_{xy}$ all take values of $\pm 1$. When we compute the Renyi entropy of a bulk region $E_{A}$ and a boundary region $\overline{A}$, the external spins are defined as
\bea
m_X=\elist{-1,&~X\in \overline{A}\\+1,&~X\notin \overline{A}},~m_{xy}=\elist{-1,&~xy\in E_A\\+1,&~xy\notin E_A}\label{bdrycondition}
\eea 
For small fluctuations around a graph $K$ considered here, we have
\bea
\overline{\trace{\rho_{A\cup b}^2}}&=&{\rm const.}\sum_{\ke{s_x=\pm 1}}e^{-\calA\kd{\ke{s_x}}}\nn\\
\calA\kd{\ke{s_x}}&=&-\frac{J_1}2\sum_{\avg{xy}\in K}\kc{s_xs_y-1}-\frac{J_2}2\sum_{\avg{xy}\notin K}\kc{s_xs_y-1}\nn\\
& &-\frac 12\log D\sum_{x\in B}s_xm_X-\frac 14\log\kc{2\Lambda +1}\sum_{xy}m_{xy}\kc{s_x+s_y}
\eea
$J_1$ and $J_2$ are the same as in Eq. (\ref{IsingactionCode}). The earlier calculation of the entropy of entire boundary in Eq. (\ref{IsingactionCode}) corresponds to the special case $m_{xy}=+1,~\forall x,y$ and $m_X=-1,~\forall X$. The constant prefactor is not important as it is the same for all configurations, and does not affect normalized quantities such as $\overline{\trace{\rho_{A\cup b}^2}}/\overline{\trace{\rho_{A\cup b}}}^2$. Similarly, $\overline{\trace{\rho_{A}}^2}$ and $\overline{\trace{\rho_b}^2}$ can be computed by the same action with different boundary conditions. 

The mutual information is determined by the correlation between external spins mediated by the dynamical spins. We denote the effective action $\calA_{\rm eff}^{--}=-\log \overline{\trace{\rho_{\overline{A}\cup E_A}^2}}$ as the effective action with boundary condition (\ref{bdrycondition}), with $--$ labeling the sign of external spin in $\bar{A}$ and $E_A$ respectively. Similarly $\calA_{\rm eff}^{-+}=-\log \overline{\trace{\rho_{\overline{A}}^2}},~\calA_{\rm eff}^{+-}=-\log \overline{\trace{\rho_{E_A}^2}}$, and $\calA_{\rm eff}^{++}=\overline{\trace{\rho}^2}$ is the normalization constant. Then 
\bea
I^{(2)}(E_A:\overline{A})=S^{(2)}_{\overline{A}}+S^{(2)}_{E_A}-S^{(2)}_{\overline{A}E_A}\simeq \calA_{\rm eff}^{+-}+\calA_{\rm eff}^{-+}-\calA_{\rm eff}^{++}
-\calA_{\rm eff}^{--}
\eea
is determined by the ``energy cost" of the external spins in $\bar{A}$ and $E_A$ being anti-parallel. The requirement of zero mutual information is equivalent to the requirement that the two external spins are completely uncorrelated. It is easy to see that this is true in the limit we consider, with $J_1, \log D\>\infty$ and $J_2, \Lambda$ finite. In this limit, the spin configuration $s_x$ is completely determined by boundary external spins $m_X$, and thus $I^{(2)}(E_A:\overline{A})=0$. For finite $J_1,\log D$, the local reconstruction condition depends on more detailed properties of the classical geometry. Although it is possible to write down some sufficient condition by taking $J_1$ and $\log D$ to be very large, we feel these conditions are not so useful to include here. 

\section{Overlap between different classical geometries}\label{sec:overlap}

In the discussion above we have shown that each classical geometry labeled by a graph $K$ is accompanied with a code subspace that satisfies bulk-boundary isometry and local reconstruction properties. The next question is whether the code subspaces for different classical geometries are truely independent subspaces of the boundary Hilbert space. Since the basis $\ket{\Psi\kd{\ke{a_{xy}}}}$ is over-complete, different geometries are generically not orthogonal, but in the following we will show that states in the code subspace of different classical geometries have exponentially small overlap. 

For this purpose we study the overlap $C_{ab}=\braket{\Psi\kd{\ke{a_{xy}}}}{\Psi\kd{\ke{b_{xy}}}}$ between two generic geometries $a_{xy}$ and $b_{xy}$. Using the definition (\ref{eq:fluctuatingRTN}) we have
\bea
C_{ab}=D^{-V_B}\trace{\prod_x\ket{V_x}\bra{V_x}\prod_{x\neq y}\ket{b_{xy}}\bra{a_{xy}}}
\eea
Carrying the random average one obtains
\bea
\overline{C_{ab}}=D^{-(V-1)V_b-VV_B}\delta_{ab}
\eea
It is essential to go to the second order and study the fluctuation around the average value, so that we evaluate $\overline{\abs{C_{ab}}^2}$:
\bea
\overline{\abs{C_{ab}}^2}&=&D^{-2V_B}\trace{\prod_x\overline{\ket{V_x}\bra{V_x}^{\otimes 2}}\kc{\prod_{x\neq y}\ket{b_{xy}}\bra{a_{xy}}\otimes \ket{a_{xy}}\bra{b_{xy}}}}\nn\\
&=&D^{-2V_B}\Omega^{-1}\sum_{R\subseteq\text{bulk}}\trace{X_R\prod_{x\neq y}\ket{b_{xy}}\bra{a_{xy}}\otimes \ket{a_{xy}}\bra{b_{xy}}}\nn\\
&=&\Omega^{-1}\sum_{R\subseteq\text{bulk}}\trace{X_R\prod_{x\neq y}\ket{b_{xy}}\bra{a_{xy}}\otimes \ket{a_{xy}}\bra{b_{xy}}}D^{-\abs{R\cap B}}\nn\\
\eea
with $\Omega=\kc{D^{V-1}+D^{2(V-1)}}^{V_b}\kc{D^{V}+D^{2V}}^{V_B}$.
To simplify this expression we can write $X_R=X_{\overline{R}}X_{\rm tot}$ with $X_{\rm tot}$ the swap of all bulk vertices. $X_{\rm tot}$ will simply permute $\ket{b_{xy}}$ and $\ket{a_{xy}}$. Relabel $\overline{R}$ by $R$ we obtain
\bea
\overline{\abs{C_{ab}}^2}&=&\Omega^{-1}\sum_{R\subseteq\text{bulk}}\trace{X_R\prod_{x\neq y}\ket{a_{xy}}\bra{a_{xy}}\otimes \ket{b_{xy}}\bra{b_{xy}}}D^{\abs{R\cap B}-V_B}\nn\\
&=&\Omega^{-1}\sum_{R\subseteq\text{bulk}}\trace{\rho_R^a\rho_R^b}D^{\abs{R\cap B}-V_B}
\eea
Here $\rho_R^{a}$ is the reduced density matrix of $\prod_{xy}\ket{a_{xy}}\bra{a_{xy}}$ in region $R$, and similarly for $\rho_R^b$. If we consider the term with $R$ the entire bulk, $\trace{\rho_R^a\rho_R^b}=|\braket{a}{b}|^2=\delta_{ab}$ is the inner-project of the two bulk states. Roughly speaking, we can consider all other terms as corrections to the overlap induced by the bulk-boundary map that is not injective. 

The overlap $\trace{\rho_{R}^a\rho_R^b}$ is nonzero only if $a_{xy}=b_{xy}$ for all $x,y\in R$. Denote the set of $R$ that satisfy this property as $\calC$. To obtain an upper bound of the overlap, we use the inequality
\bea
\trace{\rho_R^a\rho_R^b}\leq \sqrt{\trace{\rho_R^a}^2\trace{\rho_R^b}^2}=e^{-\frac12\kc{S^{(2)}_a(R)+S^{(2)}_b(R)}}
\eea
where $S^{(2)}_{a,b}(R)$ are the second Renyi entropy of states $\ket{a_{xy}}$ and $\ket{b_{xy}}$ in region $R$. Therefore
\bea
\overline{\abs{C_{ab}}^2}&\leq &\Omega^{-1}\sum_{R\in \calC}e^{-\frac12\kc{S^{(2)}_a(R)+S^{(2)}_b(R)}-\log D\kc{V_B-\abs{R\cap B}}}\label{eq:overlapbound}
\eea
To understand the physical meaning of Eq. (\ref{eq:overlapbound}), we evaluate it in several situations.
\begin{enumerate}
\item {\bf The diagonal element. } If $a_{xy}=b_{xy}~\forall x,y$, $R$ can be any subset of the bulk, and the dominant term in the sum is given by $R=\text{entire~bulk}$. Also in this case, the inequality takes the equal sign. If we take the classical geometry discussed in this section, with $J_1,\log D\>\infty$, we can ignore the contribution of other terms, and obtain $\overline{C_{aa}^2}\simeq \Omega^{-1}$. Therefore
\bea
\frac{\overline{C_{aa}^2}}{\overline{C_{aa}}^2}\simeq\kc{1+D^{1-V}}^{-V_b}\kc{1+D^{-V}}^{-V_B}
\simeq e^{-V_bD^{1-V}-V_BD^{-V}}
\eea
The ratio is close to $1$ in the limit of large volume since $V_bD^{1-V}$ and $V_BD^{-V}$ are much smaller than $1$. In other words, the fluctuation of the norm of state $\ket{\Psi\kd{\ke{a_{xy}}}}$ is exponentially suppressed, which justifies the computation of $\overline{\abs{C_{ab}}^2}$ without first normalizing the two states. 
\item {\bf Completely distinct states.} If we consider two completely distinct states such that $a_{xy}\neq b_{xy}~\forall x,y$, then the only contribution comes from $R=\emptyset$, and $\overline{\abs{C_{ab}}^2}=\Omega^{-1}D^{-V_B}=\sqrt{\overline{C_{aa}^2}\overline{C_{bb}^2}}D^{-V_B}$. In other words, the overlap between these states, after normalization, is the inverse of boundary Hilbert space dimension $D^{V_B}$. This is equal to the average overlap between two completely random states in the boundary Hilbert space dimension. \footnote{Apparently, when the bulk volume $V$ is large enough so that the basis $\ket{\Psi\kd{a_{xy}}}$ is very overcomplete, some of them will have a significant overlap. This fact, however, does not appear in the calculation of averaged overlap $\overline{\abs{C_{ab}}^2}$. The higher moments $\overline{\abs{C_{ab}}^{2k}}$ shall be able to reveal the effect of extremely large $V$, which we postpone to future works. We would like to thank Lenny Susskind for helpful discussion on this problem.}
\item {\bf Two states different in IR.} Now we study a nontrivial example. In holography all geometries considered are asymptotically anti-de Sitter space in UV (the region near the boundary) and are generically different in IR. For example we may consider two geometries, one with a black hole in IR and one without black hole. As a toy model of this situation, we can consider two geometries that are identical in a UV region $R_m$ bounding the boundary, and distinct in the IR region, as is illustrated in Fig. \ref{fig:overlap}. We assume $a_{xy}$ and $b_{xy}$ are completely distinct if $x$ or $y$ are outside region $R_m$, so that all regions contributing to the overlap are $R_m$ or its subsets. In this case the dominant contribution to Eq. (\ref{eq:overlapbound}) is given by the $R\subseteq R_m$ that has minimal averaged entropy $\frac12\kc{S^{(2)}_a(R)+S^{(2)}_b(R)}$. If both geometries are classical geometries with all connected links $a_{xy}=a_0$, the entropies satisfy area law $S^{(2)}_{a,b}(R)=s_0\abs{\pa R}_{a,b}$ with $s_0$ the entropy contributed by each link state $\ket{a_0}$. $\abs{\pa R}_{a,b}$ denotes the area (number of links crossing the boundary of $R$) in graphs of $a,b$ respectively. In summary we obtain for two classical geometries $a,b$
\bea
\frac{\overline{\abs{C_{ab}}^2}}{\sqrt{\overline{C_{aa}}^2\overline{C_{bb}}^2}}\leq e^{-\frac{s_0}2\kc{\abs{\pa R}_a+\abs{\pa R}_b}}\label{eq:overlapbound2}
\eea
where $R$ is chosen to minimize the averaged area. For example if we consider two geometries with and without a black hole, and assume that the geometry to be identical in UV until a certain distance to the horizon, then $\abs{\pa R}_{a,b}>A_{BH}$ is bounded by the area of black hole horizon, so that the overlap is upper bounded by $e^{-S_{BH}}$. More generally, the overlap is bounded by the entropy of the minimal area surface that enclose the region where the two geometries are (macroscopically) distinct. 
\end{enumerate}

\addfig{4in}{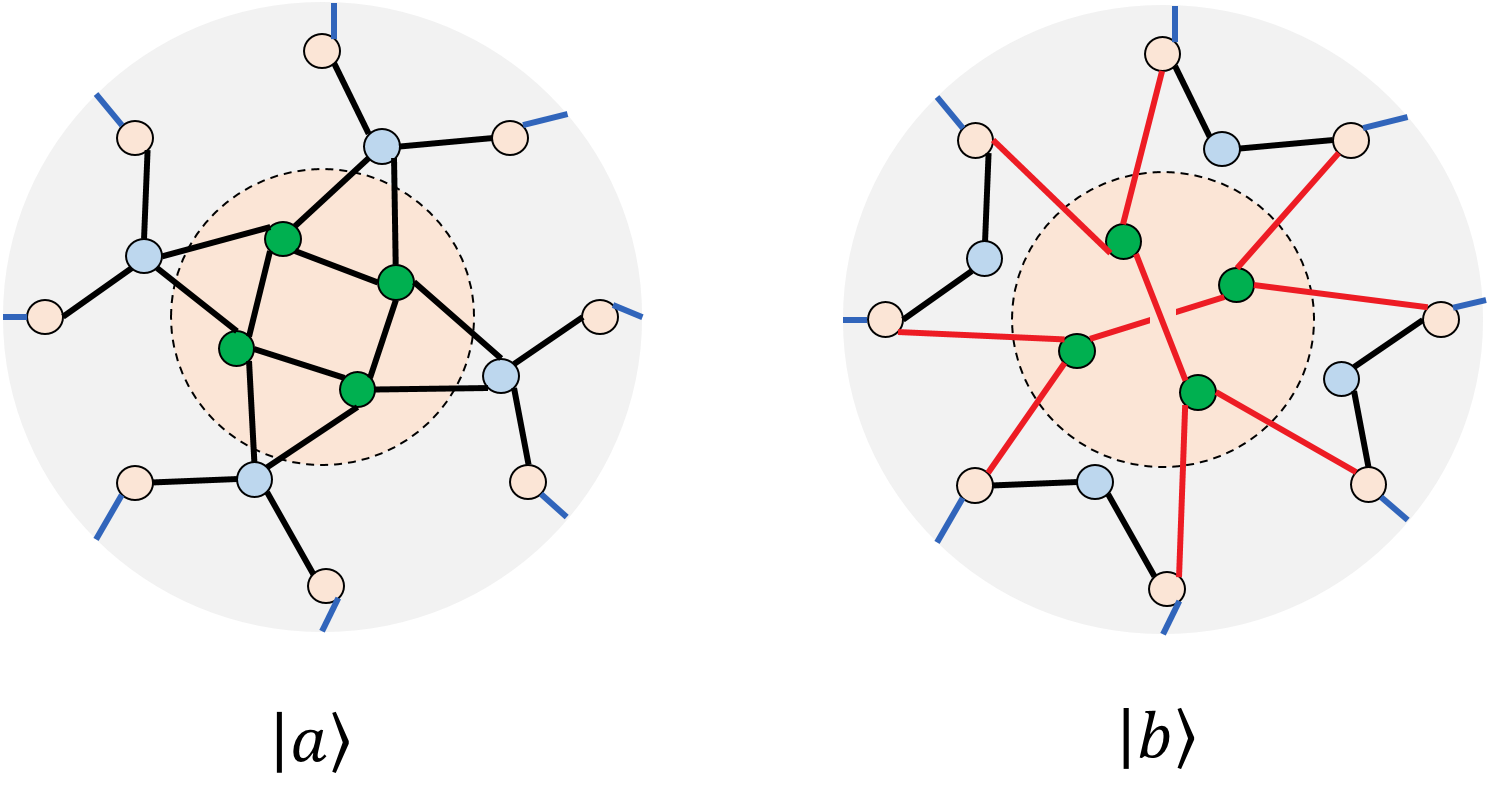}{Two graphs with identical edges in the UV region (grey) and different edges in IR (orange) and between the two regions. The overlap of these two states are upper bounded by Eq. (\ref{eq:overlapbound}), with $\abs{\pa R}_a=\abs{\pa R}_b=8$. \label{fig:overlap}}

From our definition of code subspace, it's clear that if two classical geometries are distinct at a link $xy$, the small fluctuations $a_{xy}+\delta a_{xy}$ are still distinct from $b_{xy}+\delta b_{xy}$. Therefore the overlap upper bound for $\overline{\abs{C_{ab}}^2}$ between two classical geometries $a,b$ also applies to any pair of states from the code subspaces of $a$ and $b$. Consequently, if we choose a set of macroscopically distinct geometries $a_{xy}^n$, the code subspaces $\mathbb{H}_{Cn}$ of each of them are almost orthogonal subspaces of the boundary Hilbert space. One can define a bigger code subspace $\mathbb{H}_C=\oplus_n\mathbb{H}_{Cn}$ such that the bulk-boundary isometry is still well-defined in the bigger code subspace. In the bigger code subspace $\mathbb{H}_C$, operators that can be reconstructed on a boundary region $A$ form an algebra with nontrivial center, a structure that has been investigated in Ref. \cite{harlow2016ryu}. We would like to comment a bit more on the mapping between bulk and boundary operators. A generic bulk operator in this code subspace has the form $\phi=\sum_n P_n \phi_n P_n$, with $P_n$ the projection operator onto $n$-th code subspace $\mathbb{H}_{Cn}$, and $\phi_n$ an operator acting only in that subspace. If we denote the linera map from boundary to bulk as $M$, a local operator $\phi_n$ in the code subspace of geometry $a^n_{xy}$ is mapped to a boundary operator $M^\dagger P_n \phi_n P_nM$. Although the bulk-boundary mapping is linear and isometric, one can consider $P_nM$ as the linear map restricted to a code subspace, which is ``state-dependent"\cite{papadodimas2013infalling}. Locality in the bulk can only be defined in a code subspace around a given classical geometry, and the local operators in a code subspace (such as an operator $\phi_{xy}$ that only slightly changes $a_{xy}$ value for one link) is actually an operator $P_n \phi_{xy} P_n$ in the large bulk Hilbert space. The ``state dependence" of operator correspondence in each code subspace is encoded in the support of the operator in the bulk Hilbert space, specified by $P_n$. 

\section{Conclusion and discussions}\label{sec:conclusion}

In conclusion, we have shown that the random tensor network states on all graphs form an overcomplete basis of the boundary Hilbert space, which we name as holographic coherent states. A generic boundary state is mapped to a superposition of geometries. The semiclassical geometries are defined as small fluctuations around reference classical geometries with strongly entangled edges. We show that small fluctuations around a classical geometry form a code subspace, the states in which are mapped to the boundary isometrically, with local reconstruction properties. Furthermore, we show that states in the code subspaces of two different classical geometries are almost orthgonal to each other, with their overlap decaying exponentially as a function of the minimal area surface that covers the bulk region in which the two bulk geometries are distinct. 

The holographic coherent state basis has a lot of similarity to the
coherent state basis of a boson field. If we consider a complex boson
field described by a $\abs{\phi}^4$ theory, the coherent state basis
$\ket{\phi(x)}$ is an overcomplete basis of the system, with which one
can write a path integral representation of the partition
function. The action of the system may have multiple local minima, for
example configurations with and without vortices. Around each local
minimum one can expand the action in small fluctuations,
$\calA\kd{\phi_c+\delta \phi}\simeq
\calA_c+\frac12\frac{\delta^2\calA}{\delta\phi\delta\phi}\delta\phi\delta\phi$. The
quantization of such fluctuations are low energy quasiparticles such
as superfluid phonons. The Hilbert space of such quasiparticle
excitations is a ``low energy subspace" of the entire Hilbert
space. Different classical minima
$\ket{\phi_{c1}(x)},~\ket{\phi_{c2}(x)}$ are not exactly orthogonal,
but the overlap of macroscopically different states are exponentially
suppressed. Therefore one can view the low energy excitations
associated with each of them as physically independent
subspaces. \footnote{It is interesting to note that the $\phi^4$
  theory example appeared in a related discussion in
  Ref. \cite{harlow2014aspects} about state-dependent operators (see
  Sec. 5).} There are two key differences between the holographic
coherent states we consider and the boson coherent states. Firstly,
the overlap in the former case is suppressed by exponential of the
minimal area covering the distinct region, while that in the latter
case is suppressed by exponential of the volume of the distinct
region, which can be viewed as a manifestation of holographic
principle. Secondly, in the gravity case, locality in the bulk is only
defined in the code subspaces, which can be seen in the fact that the
log of Hilbert space dimension $\log({\rm dim}(\mathbb{H}_C))$ is
proportional to the volume of the bulk, while that of the total
Hilbert space $\log({\rm dim}(\mathbb{H}))$ is proportional to the
boundary. On comparison, in ordinary boson coherent state case both
quantities are proportional to the volume of the system. 

There are a lot of open questions along this direction. For a given boundary Hamiltonian, a natural problem is to use the holographic coherent states as variational wavefunctions. The geometry described by $a_{xy}$ can be used as a ``mean-field order parameter" that is optimized by minimizing the energy. The difficulty of this approach is the random average, which introduces the ambiguity of a local unitary transformation and therefore mixes states with very different energy. In principle, this problem can be solved in the following procedure. For each given geometrical state $\ket{a}\equiv \ket{\Psi\kd{\ke{a_{xy}}}}$, one can consider all local unitary transformations $\prod_{X\in B}^\otimes u_X\ket{a}$, with $u_X\in SU(D)$, and variationally determine $u_X$ by minimizing energy. Denote the minimal energy in this class of states as $E\kd{a}$, we can then minimize energy to determine the optimal bulk geometry $a_{xy}$. It is not clear whether such a variational procedure is technically feasible. We will reserve that to future works. 

Another natural question is how to obtain the bulk equation of motion---the analog of Einstein's equation. By writing the boundary dynamics into a path integral in the geometrical basis, one can in principle obtain a bulk action. Is the Einstein equation or its analog the saddle point equation if the bulk action? Will such saddle point equation be related to previous entanglement approaches to Einstein equation\cite{lashkari2014gravitational,faulkner2014gravitation,swingle2014universality,
jacobson2016entanglement,lloyd2012quantum,verlinde2011origin}. Yet another interesting question is whether a similar area-law bound of state inner product exists in general relativity, where the inner product between two states is defined by a path integral with these states as boundary conditions\cite{hartle1983wave,maldacena2003eternal}.  It is interesting to compare our results with other recent discussions about the overcompleteness of the geometry basis.\cite{jafferis2017bulk,nomura2016toward,nomura2016spacetime}

\noindent{\bf Acknowledgement.} We would like to acknowledge helpful discussions with Ahmed Almheiri, Patrick Hayden, Aitor Lewkowycz, Don Marolf, Sepehr Nezami, Leonard Susskind and Michael Walter. This work is supported by the National Science Foundation through
the grant No. DMR-1151786 (ZY), and the David and Lucile Packard Foundation (XLQ).

\appendix

\section{Fluctuations and higher Renyi entropies}\label{app:fluctuation}


In Sec.\ref{sec:isometry}, we make the following approximation in the calculation of the second Renyi entropy. 
\begin{equation}
  \label{eq:1}
  \overline{e^{-S_B^{(2)}}} =  \overline{\left(\frac{\text{Tr}[\rho_B^2]}{\text{Tr}[\rho_B]^2}\right)} \approx \frac{e^{-A^{(2)}_{min}[h_{1}]}}{e^{-A^{(2)}_{min}[h_{0}]}}
\end{equation}
where $h_1$, $h_0$ denote the boundary field configuration for the
calculation of $\text{Tr}[\rho_B^2]$, $\text{Tr}[\rho_B]^2$.  This
calculation is valid if the fluctuation around the minimum is
small \cite{hayden2016holographic}. Formally, the following conditions
should be satisfied
$\overline{\left(\frac{\text{Tr}[\rho_B^2]}
    {e^{-A^{(2)}_{min}[h_{1}]}} - 1 \right)^2}\ll 1$, which can be
achieved by requiring
\begin{equation}
\label{eq:fluct first}
      \overline{\left( \frac { \text{Tr}[\rho_B^2] } {e^{-A^{(2)}_{min}[h_{1}]}} - 1 \right)^2}
    \leq \frac {\overline{\text{Tr}[\rho_B^2]^2}} {e^{-A^{(4)}_{min}[h_{1}]}} - 1 \ll 1
\end{equation}
Here we have used that
$\overline{\text{Tr}[\rho_B^2]} \geq e^{-A^{(2)}_{min}[h_{1}]}$, since
at finite temperature the partition function receives contributions
from all spin configurations, not just the minimal energy
configuration. Similarly for the calculation of $\trace{\rho_B}^2$ one can require 
$\frac {\overline{\text{Tr}[\rho_B]^4}} {e^{-A^{(4)}_{min}[h_{0}]}} - 1 \ll 1$.
Thus the calculation of the fluctuation requires the random average over four copies of the density
matrix. Similarly, when calculating the $k$th Renyi entropy,
we need to calculate $\trace{\rho_B^k}$ which involves $k$ copies of the density matrix. For example $\overline{\trace{\rho_B^4}}$ and $\overline{\trace{\rho_B^2}^2}$ are both average of $4$ copies of density matrices, with different boundary conditions which specify the contraction of indices. More explicitly they can be written as $\overline{\trace{\rho_B^4}}=\overline{\trace{\rho^{\otimes 4}h_{(1234)}^B}}$ and $\overline{\trace{\rho_B^2}^2}=\overline{\trace{\rho^{\otimes 4}h_{(12)(34)}^B}}$ with $h_{(1234)}^B$ the cyclic permutation acting on $4$-copies of $B$, and $h_{(12)(34)}^B$ the permutation of $12$ and $34$ acting on the same region. 

Therefore in general we can evaluate the random average of $k$ copies of density matrix with an arbitrary
boundary condition, and study how to control its deviation from the contribution of the dominant configuration. The $k$ copy quantity with most general boundary condition can be expressed as
\bea
\overline{Z^{(k)}}&\equiv &\trace{\overline{\rho^{\otimes k}}\prod_{X\in B}h_X}\nn\\
&=&\sum_{g_x^i\in S_k}\prod_{xy}\trace{\rho_{xy}^{\otimes k}g_x^ig_y^j}\prod_{x\in B}\trace{\rho_{EPR}^{\otimes k}g_x^ih_X}\label{eq:partitionfunction}
\eea
with boundary permutations $h_X\in S_k$ defining the boundary conditions. We label the permutation group elements as $g_x^i,~i=0,1,2,...,k!-1$, with $g_x^0=I_x$ the identity operator. The averaged entanglement quantity is mapped to a partition function of a $S_k$ statistical mechanical model defined on the complete graph.

In the following we will prove that the fluctuation of such quantities with general boundary conditions is bounded if the following sufficient conditions are satisfied:
\begin{eqnarray}
  &&(V-1)\log D_L \gg 2\log D \label{eq:sufficient} \\
&& \text{Tr}(\rho_{xy}^{\otimes k}g_x^ig_y^i)\text{Tr}(\rho_{xy}^{\otimes k}g_x^jg_y^j)  > |\text{Tr}(\rho_{xy}^{\otimes k}g_x^ig_y^j)|^2 , ~~\forall, i\neq j \label{eq:suffL}  
\end{eqnarray}

In Sec.\ref{GR}, we bound the fluctuations based on conditions \ref{eq:sufficient} and
Eq.\ref{eq:suffL}. In Sec.\ref{app:rhoxycondition}, we propose a
stronger condition of the density matrix that implies
Eq.\ref{eq:suffL}. In Sec.\ref{app:spinexample}, we construct the an
explicit example in spin system and show that Eq.\ref{eq:sufficient}
and Eq.\ref{eq:suffL} are satisfied.

\subsection{General results}\label{GR}

In this section, we prove that Eq.\ref{eq:sufficient} and
Eq.\ref{eq:suffL} are sufficient to bound the fluctuations and to
guarantee that higher Renyi entropies are close to the maximum. 

First we rewrite Eq.\ref{eq:suffL} as 
\begin{equation}
  \label{eq:2}
  L^{ii}(k) + L^{jj}(k)  - 2L^{ij}(k) > 0
\end{equation}
where
\begin{equation}
  L^{ij}(k) = -\frac{1}{2}\left(\log\text{Tr}(\rho_{xy}^{\otimes k}g_x^iI_y) + \log\text{Tr}(\rho_{xy}^{\otimes k}I_xg_y^j) - \log\text{Tr}(\rho_{xy}^{\otimes k}g_x^ig_y^j) \right)
\end{equation}

Next, it is straightforward to show that Eq.\ref{eq:suffL} implies
$L^{ii}(k)>0$, $i\neq 0$, a condition we will use to bound the fluctuation. If we
take $j=0$, Eq.\ref{eq:suffL} means
\begin{equation}
  \label{eq:6}
  L^{ii}(k) + L^{00}(k)-2L^{i0}(k) >0
\end{equation}
Since $\rho_{xy}$ is normalized. $L^{00}(k)=L^{i0}(k)=0$. Thus $L^{ii}(k)>0$.

Now we calculate the partition function in ($k>2$) replica with an arbitrary boundary condition. Using permutation symmetry between vertices in the complete graph, Eq. \ref{eq:partitionfunction} can be rewritten as
\begin{eqnarray}
  && \overline{Z_1^{(k)}} = \sum_{\{n_i\},\{m_i\} }e^{-A(n_i,m_i)}\frac{V_b!}{n_0!n_1!\cdots n_{k!-1}!}\frac{V_B!}{m_0!m_1!\cdots m_{k!-1}!} \\
\nonumber  && A(n_i,m_i)= \sum_{i> j} J^{ij}\left(n_i+m_i\right)\left(n_j+m_j\right) + \sum_i \frac{J^{ii}}{2}(n_i+m_i)(n_i+m_i-1) + \sum_{i} B^i m_i \\
\nonumber  \text{with~}&& J^{ij} = -\log \text{tr} \left(\rho_{xy}^{\otimes k} g_x^i g_y^j \right) ~~~~
     B^i = -\log \text{tr}\left(\rho_{\text{EPR}}^{\otimes k} g_x^i h  \right) ~~~~ L^{ij} = (J^{i0}+J^{0j}-J^{ij})/2\\
\nonumber  && \sum_{i} n_i = V_b ~~~~ \sum_i m_i = V_B
\end{eqnarray}
where $n_i(m_i)$ is the number of bulk(boundary) points occupied by
the group element $g^i$; $V_b(V_B)$ is the total number of
bulk(boundary) points; $h_X=h$ fixes the boundary condition.

Then we replace $n_0 = V_b - \sum_{i\geq 1} n_i$,
$m_0 = V_B - \sum_{i\geq 1} m_i$. Since $J^{00}=0$, we have
\begin{eqnarray}
\nonumber  A(n_i,m_i)
  &=& \sum_{ i> j\geq 1} J^{ij}\left(n_i+m_i\right)\left(n_j+m_j\right) + \sum_{i\geq 1}J^{i 0} \left(n_i+m_i\right)\left( V_b + V_B - \sum_{j\geq 1}(n_j + m_j)\right)\\
\nonumber  && + \sum_{i\geq 1} \frac{J^{ii}}{2}(n_i+m_i)(n_i+m_i-1) 
     + \sum_{i} B^i m_i \\
 \nonumber &=& \sum_{i,j\geq 1} -(n_i+m_i)L^{ij}(n_j+m_j) + \sum_{i\geq 1} \left((V_b+V_B)L^{ii}+(V_b+V_B-1)\frac{J^{ii}}{2}\right)n_i \\
  &&+\sum_{i\geq 1} \left((V_b+V_B)L^{ii}+(V_b+V_B-1)\frac{J^{ii}}{2}+B^i-B^0\right)m_i + B^0 V_B
\end{eqnarray}

In the large $V_b,V_B$ limit, we treat $n_i$ and $m_i$ as continuous variables to decide where
$F(n_i,m_i) \equiv - A(n_i,m_i) - \sum_i\log n_i!- \sum_i\log m_i!$
reaches its maximum. We use Stirling formula and calculate the second
derivatives of this function
\begin{eqnarray}
  M &=& \begin{bmatrix}
    M_1       &  M_2 \\
    M_2    &  M_3 
\end{bmatrix} \\
\nonumber   M_1^{ij} &=& \frac{\partial^2}{\partial n_i\partial n_j} F(n_i,m_j) = 2L^{ij} - \frac{\delta_{ij}}{n_i} - \frac{1}{n_0} \\
\nonumber  M_2^{ij} &=&\frac{\partial^2}{\partial n_i\partial m_j} F(n_i,m_j) = 2L^{ij} \\
\nonumber  M_3^{ij} &=& \frac{\partial^2}{\partial m_i\partial m_j} F(n_i,m_j) = 2L^{ij} - \frac{\delta_{ij}}{m_i} - \frac{1}{m_0} 
\end{eqnarray}
Now we show that $F$ does not have local minimum away from the corners of the parameter space. A local minimum requires $M$ to be a negative definite matrix, so to prove that $F$ does not have local minimum one just needs to show that $M$ is not negative definite anywhere away from the corners. A corner of the parameter space (labeled by $n_i/V_b,~m_i/V_B$) is defined by having one $n_i=V_b,~m_j=V_B$ and all other numbers vanishing. Therefore for any point away from these corners, there are either two numbers $n_i,~n_j$ of order $V_b$, or two numbers $m_i,~m_j$ of order $V_B$. Let's assume there are $n_i,~n_j$ of order $V_b$ since the discussion with $m_i,~m_j$ is exactly in parallel. This includes the following two cases:

\begin{itemize}
\item If $n_0$ is of $O(1)$, then there are two $n_i$,
  $n_j$ with $i,j>0$ of order $O(V_b)$. Define a vector $\vec{v}$ whose
  $i$th element is $1$, $j$th element is $-1$ and all others are
  $0$. Obviously,
  \begin{equation}
    v^T M v = 2 \left(L^{ii} +L^{jj}-2L^{ij}\right) - \frac{1}{n_i} - \frac{1}{n_j}
  \end{equation}
  Since $L^{ii} +L^{jj}-2L^{ij}>0$, and $n_i$, $n_j$ are $O(V_b)$,
  $v^T M v>0$. So $M$ is not negative definite and there is no local
  maximum in this case away from the corners.
\item If $n_0$ is of $O(V_b)$, then there is at least another $n_i$
  being $O(V_b)$. We choose $\vec{v}$ whose only non-zero element is
  $1$ at the $i$th element. Thus
  \begin{equation}
    \label{eq:3}
     v^T M v = 2 L^{ii} - \frac{1}{n_i} - \frac{1}{n_0}
  \end{equation}
  Since $L^{ii}>0$ is $O(1)$ and $n_0$, $n_i$ is $O(V_b)$,
  $v^T M v>0$.
\end{itemize}
Therefore we conclude that Eq.\ref{eq:suffL} is the sufficient condition that guarantees $F(n_i,m_i)$ does not have local minimum away from the corners.

The next step is to compare the value of $F(n_i,m_i)$ of each corner solution and bound the near corner solutions. The corner solutions are categorized as 
\begin{itemize}
\item $S_{n_0,m_0}$: $n_0=V_b$, $m_0=V_B$, 
  \begin{equation}
    F(S_{n_0,m_0})= -B^0 V_B
  \end{equation}
\item $S_{n_i,m_0}$: $n_i=V_b$, $i\geq 1$, $m_0=V_B$, 
  \begin{eqnarray}
    F(S_{n_i,m_0})=  -B^0 V_B - V_bV_B L^{ii} - (V-1)\frac{J^{ii}V_b}{2}
  \end{eqnarray}
\item $S_{n_0,m_j}$: $n_0=V_b$, $m_j=V_B$,  $j\geq 1$,
  \begin{eqnarray}
    F(S_{n_0,m_j}) = -B^j V_B - V_bV_B L^{jj} - (V-1)\frac{J^{jj}V_B}{2}
  \end{eqnarray} 
\item $S_{n_i,m_j}$: $n_i=V_b$, $m_j=V_B$,  $i,j\geq 1$, 
  \begin{eqnarray}
    F(S_{n_i,m_j}) = -B^j V_B -V_bV_B \left(L^{ii}+L^{jj}-2L^{ij}\right)
        - (V-1)\frac{J^{ii}V_b+J^{jj}V_B}{2}
  \end{eqnarray} 
\end{itemize}

Firstly, we notice that $ F(S_{n_0,m_0}) \gg F(S_{n_i,m_0})$ is always
true, because $L^{ii}>0$ is assumed and $J^{ii}= \log D_L \left(k - \chi(g^i)\right)>0$, where
$\chi(g)$ denotes the number of cycles in a permutation
$g$.

Secondly,
\begin{eqnarray}
  \nonumber &&F(S_{n_0,m_0})-F(S_{n_0,m_j})\\
 \nonumber &=& V_B\log D\left(k - \chi((g^{j})^{-1} h) - (k - \chi( h)) \right)+ V_B V_b L^{ii}  + (V-1)\frac{J^{jj}V_B}{2} \\
 &\geq& - V_B\log D\left(k - \chi((g^{j})^{-1}) \right) +  (V-1)\frac{V_B}{2} \log D_L \left(k-\chi((g^{j})^{-1}) \right)
\end{eqnarray}
In the inequality, we use $L^{ii}>0$, and the triangle inequality of
$d(g,h)\equiv k -\chi(g^{-1}h) $, which is equal to the minimal number
of transpositions (i.e., permutations that exchange only two indices)
required to write a permutation $g^{-1}h$. $d(g,h)$ defines a distance on $S_k$, which satisfies the triangle inequality $d(g,I)+d(I,h)\geq d(g,h)$\cite{hayden2016holographic}. Thus Eq.\ref{eq:sufficient}  $ (V-1)\log D_L \gg 2\log D $
is a sufficient condition for $ F(S_{n_0,m_0}) \gg F(S_{n_0,m_j})$.

Thirdly,
\begin{eqnarray}
  \nonumber &&F(S_{n_0,m_0})-F(S_{n_i,m_j})\\
 \nonumber &=& V_B\log D\left(k - \chi((g^{j})^{-1} g^\partial) - (k - \chi( g^\partial)) \right)+ V_B V_b (L^{ii}+L^{jj}-2L^{ij})  \\
 \nonumber   &&+ (V-1)\frac{J^{ii}V_b+J^{jj}V_B}{2} \\
\nonumber &>&  V_B\log D\left(k - \chi((g^{j})^{-1} g^\partial) - (k - \chi( g^\partial)) \right) +  (V-1)\frac{V_B}{2} \log D_L \left(k-\chi((g^{j})^{-1}) \right) \\
 &\geq& - V_B\log D\left(k - \chi((g^{j})^{-1}) \right) +  (V-1)\frac{V_B}{2} \log D_L \left(k-\chi((g^{j})^{-1}) \right)
\end{eqnarray}
where in the first inequality, we use $J^{ii}>0$ and  $L^{ii}+L^{jj}-2L^{ij}>0$. In the second inequality, we use the
triangle inequality of $k -\chi((g^i)^{-1}g^j) $ again. Thus if
Eq.\ref{eq:sufficient} holds, we also have $F(S_{n_0,m_0})\gg F(S_{n_i,m_j})$.
 
In fact, we can make tighter bounds in
$F(S_{n_0,m_0})-F(S_{n_i,m_j})$ and
$F(S_{n_0,m_0})-F(S_{n_0,m_j})$ if we do not simply discard
$L^{ii}$ or $L^{ii}+L^{jj}-2L^{ij}$. However, using condition
Eq.\ref{eq:sufficient} has the advantage that it does not depend on $k$ and the details of the link state.

Finally, we can bound $\overline{Z_i^{(k)}}$ by analyzing the
configurations near the corners. We have shown that when
Eq.\ref{eq:sufficient} and Eq.\ref{eq:suffL} are satisfied, all other
corner solutions are exponentially small compared with the dominating
corner $S_{n_0,m_0}$, and the exponent is suppressed by
$-\frac{k-\chi(g^j)}{2} V_B \left((V-1)\log D_L - 2\log D\right)
$. Thus the next biggest configuration is at the neighborhood of the
corner solution $S_{n_0,m_0}$. In fact we can bound all configurations
that are finite distance away from $S_{n_0,m_0}$ by
$C\cdot \exp\left[- \frac{(V-1)\log D_L -2\log D}{2}\right]$, where
$C$ is a $O(1)$ number. Thus we obtain that
\begin{equation}
  \label{eq:4}
  \overline{Z_i^{(k)}} \leq e^{- B^0 V_B}\left(1 + C (V_B V_b)^{k!-1}  \exp\left[- \frac{(V-1)\log D_L -2\log D}{2}\right]\right)
\end{equation}
where $(V_B V_b)^{k!-1} $ is the total number of configurations of $F(n_i,m_j)$.

We conclude that if Eq.\ref{eq:sufficient}, \ref{eq:suffL} are
satisfied, the fluctuation is controlled and all higher Renyi entropies are close to $V_B \log D$. Thus there is an isometry from the
boundary to the bulk.

\subsection{A sufficient condition for Eq.\ref{eq:suffL}} \label{app:rhoxycondition}

In this section, we provide a sufficient condition that deduces Eq.\ref{eq:suffL}, which helps to clarify what density matrices satisfy this equation. In a basis $\ket{\alpha_x}=\prod_{s=1}^k\ket{\alpha_{x}^k}$ of the $k$-copied Hilbert space, density operators and permutation operators are written as
\begin{eqnarray}
  \rho_{xy}^{\otimes k} &=&  \left(\rho_{xy}^{\otimes k} \right)_{\alpha,\beta,\gamma,\delta} \left(|\alpha_x\rangle\otimes |\beta_y\rangle\right) \left(\langle \gamma_x|\otimes \langle \delta_y|\right) \\
g^{j} &=& (g^j)_{\alpha,\beta} |\alpha\rangle \langle\beta|
\end{eqnarray}
One can rearrage the indices and write
\begin{eqnarray}
&&  \text{Tr}\left[\rho_{xy}^{\otimes k} g_x^i g_y^j \right] = (g^i)_{\gamma,\alpha}   \left(\rho_{xy}^{\otimes k}  \right)_{\alpha,\beta,\gamma,\delta} (g^j)_{\delta,\beta}  = \langle g^i| \tilde{\rho}_{xy}^{\otimes k} | g^j\rangle \\
\label{rhoxyt}&&\tilde{\rho}_{xy}^{\otimes k} \equiv  \left(\rho_{xy}^{\otimes k}  \right)_{\alpha,\beta,\gamma,\delta}  \left(|\alpha_x\rangle\otimes |\gamma_x\rangle\right)\left(\langle \beta_y|\otimes \langle \delta_y|\right) \\
&& | g^j\rangle  \equiv (g^j)_{\alpha,\beta} |\alpha\rangle\otimes |\beta\rangle, ~~
\langle g^i |  \equiv (g^i)_{\alpha,\beta} \langle \alpha|\otimes \langle\beta|
\end{eqnarray}
where in the last step, we have used the fact that the matrix elements $g^i_{\alpha,\beta}$ in the product basis are real. In this representation, $\text{Tr}\left[\rho_{xy}^{\otimes k} g_x^i g_y^j \right] $ becomes an inner product between states $\ket{g^i},\ket{g^j}$ with metric $\tilde{\rho}_{xy}^{\otimes k}$. Therefore Eq.\ref{eq:suffL} follows from Cauchy-Schwarz inequality if $\tilde{\rho}_{xy}^{\otimes k}$ is Hermitian and
positive semi-definite for all $k$.  Thus we conclude that a sufficient but
not necessary condition for Eq.\ref{eq:suffL} is that $\tilde{\rho}_{xy}$ is Hermitian and positive semi-definite. This condition is not necessary since Eq. (\ref{eq:suffL}) is only required for permutation operators and does not need to hold for general operators. 

\subsection{An explicit example of states $\ket{a_{xy}}$}\label{app:spinexample}


In this section, we provide an explicit example of link states $\ket{a_{xy}}$ and prove that condition (Eq.\ref{sufficientcond2}) is satisfied. We define the state $\ket{J}$ as a SU(2) singlet formed by two spins each carrying spin $J$ representation:
\bea
 |J\rangle &\equiv&  \sum_{M} \frac{(-)^{J-M}}{\sqrt{2J+1}} |J, M;J, -M \rangle
\eea
with $J=0,1,...,D_L-1$ labeling the link states. The Hilbert space of each site is a direct sum of different representations $\mathbb{H}_x=\oplus_{J=0}^{D_L-1}\mathbb{H}_J$. States with different $J$ obviously are orthogonal. (A subtlety is that the entropy of state $\ket{J}$ is $\log\kc{2J+1}$, so that we should think the link variable $a\propto\log\kc{2J+1}$ if we still want $a$ to label the entropy across the link. This does not affect any discussion here.) The density matrix $\rho_{xy}$ is given by
\begin{eqnarray}
\label{DM}  \rho_{xy} &\equiv& \frac{1}{D_L-1}\sum_{J=1}^{D_L} |J\rangle\langle J| 
\end{eqnarray}

If one directly obtains $\tilde{\rho}_{xy}$ in Eq. (\ref{rhoxyt}) for $\rho_{xy}$, the resulting $\rho_{xy}$ is Hermitian but not positive semi-definite. However, we can prove $\rho_{xy}$ satisfies condition (\ref{eq:suffL}) by defining a unitary operator on the $y$ site
\bea
u_y=\sum_{J,M}(-1)^{M}\ket{J,-M}\bra{J,M}
\eea
The density matrix in the new basis is
\bea
\sigma_{xy}\equiv u_y\rho_{xy}u_y^\dagger=\frac1{D_L-1}\sum_{J=0}^{D_L-1}\frac1{2J+1}\sum_{M,N}\ket{J,M;J,M}\bra{J,N;J,N}
\eea
Since $u_y^{\otimes k}$ commutes with permutation operators $g_y^i$, we have $\trace{\sigma_{xy}^{\otimes k}g_x^ig_y^j}=\trace{\rho_{xy}^{\otimes k}g_x^ig_y^j}$. For $\sigma_{xy}$, the corresponding operator $\tilde{\sigma}_{xy}$ defined in Eq. (\ref{rhoxyt}) is
\bea
\tilde{\sigma}_{xy}=\frac1{D_L-1}\sum_{J=0}^{D_L-1}\frac1{2J+1}\sum_{M,N}\ket{J,M;J,N}\bra{J,M;J,N}=\oplus_{J=0}^{D_L-1}\frac1{(D_L-1)(2J+1)}\mathbb{I}_{J}
\eea
with $\mathbb{I}_J$ an identity matrix of the size $(2J+1)^2\times (2J+1)^2$. Obviously $\tilde{\sigma}_{xy}$ is diagonal and positive definite, so that we prove $\sigma_{xy}$ and therefore $\rho_{xy}$ satisfy Eq. (\ref{eq:suffL}). 

\bibliography{refs}
\end{document}

%% file: XLdef.tex
\def\text#1{{\rm #1}}

\def\avg#1{\left\langle#1\right\rangle}
\def\bra#1{\left\langle#1\right|}
\def\ket#1{\left|#1\right\rangle}
\def\braket#1#2{\left\langle #1\right|\left.#2\right\rangle}

\def\abs#1{\left|#1\right|}
\def\kc#1{\left(#1\right)}
\def\kd#1{\left[#1\right]}
\def\ke#1{\left\{#1\right\}}

\def\trace#1{{\rm Tr}\left[#1\right]}

\def\be{\begin{equation}}       \def\ee{\end{equation}}
\def\bea{\begin{eqnarray}}      \def\eea{\end{eqnarray}}
\def\ba{\begin{array} }
\def\ea{\end{array} }

\def\nn{\nonumber}
\def\pa{\partial}
\def\=>{\Rightarrow}
\def\>{\rightarrow}

\def\vect#1{\kc{\ba{c}#1\ea}}
\def\elist#1{\left\{\ba{cc} #1\ea\right.}

\def\addfig#1#2#3{\begin{figure}
\centerline{\includegraphics[width=#1]{#2}}
\caption{#3
}
\end{figure}}